\documentclass[journal=jpcbfk,manuscript=article]{achemso}
\usepackage[utf8]{inputenc}

\usepackage{longtable}
\usepackage{multirow}
\usepackage{amsmath}
\usepackage[usenames,dvipsnames]{color}
\usepackage{soul}
\usepackage{threeparttable}
\usepackage{xr}
\usepackage{graphicx} \usepackage{amsmath,amssymb}
\usepackage{caption} \usepackage{color} \usepackage{xcolor}
\usepackage{dcolumn} \usepackage{bm} \usepackage{float}
\usepackage{subcaption} \usepackage{chemformula}
\usepackage{textcomp} \usepackage{url} \usepackage[normalem]{ulem}
\usepackage{comment}

 \usepackage{microtype}
\usepackage{etoolbox}

\usepackage[unicode=true, bookmarks=true, bookmarksnumbered=true,
  bookmarksopen=true, bookmarksopenlevel=2, breaklinks=false,
  pdfborder={0 0 1}, backref=false, colorlinks=true, hidelinks
]{hyperref}

\SectionNumbersOn

\author{Cangtao Yin} \affiliation[University of
  Basel]{Department of Chemistry, University of Basel,
  Klingelbergstrasse 80, CH-4056 Basel, Switzerland.}
  
\author{Silvan K\"aser} \affiliation[University of Basel]{Department of
  Chemistry, University of Basel, Klingelbergstrasse 80, CH-4056
  Basel, Switzerland.}\altaffiliation{Present Address: Roche Pharma
  Research and Early Development, Pharmaceutical Sciences, Roche
  Innovation Center Basel, F. Hoffmann-La Roche Ltd, Basel,
  Switzerland}

\author{Meenu Upadhyay} \affiliation[University of
  Basel]{Department of Chemistry, University of Basel,
  Klingelbergstrasse 80, CH-4056 Basel, Switzerland.}
  
\author{Markus Meuwly} \affiliation[University of Basel]{Department of
  Chemistry, University of Basel, Klingelbergstrasse 80, CH-4056
  Basel, Switzerland.}  \email{m.meuwly@unibas.ch}
  
\title{End-to-End Photodissociation Dynamics of Energized H$_2$COO}

\begin{document}
\date{\today}

\begin{abstract}
The end-to-end dynamics of the smallest energized Criegee
intermediate, H$_2$COO, was characterized for vibrational excitation
close to and a few kcal/mol above the barrier for hydrogen
transfer. From an aggregate of at least 5 $\mu$s of molecular dynamics
simulations using a neural network-representation of
CASPT2/aug-cc-pVTZ reference data, the branching ratios into molecular
products HCO+OH, CO$_2$+H$_2$, or H$_2$O+CO was quantitatively
determined. Consistent with earlier calculations and recent
experiments, decay into HCO+OH was found to be rare $(\sim 2 \%)$
whereas the other two molecular product channels are accessed with
fractions of $\sim 30 \%$ and $\sim 20 \%$, respectively. On the 1 ns
time scale, which was the length of an individual MD simulation, more
than 40 \% of the systems remain in the reactant state due to partial
intramolecular vibrational redistribution (IVR). Formation of
CO$_2$+H$_2$ occurs through a bifurcating pathway, one of which passes
through formic acid whereas the more probable route connects the
di-radical OCH$_2$O with the product through a low-lying transition
state. Notably, none of the intermediates along the pathway accumulate
and their maximum concentration always remains well below 5 \%. This
work demonstrates that atomistic simulations with global reactive
machine-learned energy functions provide a quantitative understanding
of the chemistry and reaction dynamics for atmospheric reactions in
the gas phase.
\end{abstract}

\section{Introduction}
The hydroxyl radical (OH) is one of the most effective oxidizing
agents and plays an essential role in the chemical evolution of the
atmosphere.\cite{stone:2012} Given its central role in degrading
reactions of a large number of pollutants such as volatile organic
compounds and as a chain initiator in many oxidation processes, OH is
also being referred to as the ``detergent of the
troposphere''.\cite{gligorovski2015environmental,levy1971normal} For
chemical models of the lower atmosphere the amount of OH generated
from alkene ozonolysis is an important determinant. Alkene ozonolysis
proceeds through a 1,3-cycloaddition of ozone across the C=C bond to
form a primary ozonide which then decomposes into carbonyl compounds
and energized carbonyl oxides, known as Criegee Intermediates
(CIs)\cite{criegee1949ozonisierung}. Such high-energy intermediates
undergo either rapid unimolecular decay to OH\cite{alam2011total} or
stabilize through collisions.\cite{novelli2014direct} Stabilized CIs
can isomerize and dissociate into molecular products including the OH,
or undergo bimolecular reactions with other molecular species such as
water, SO$_2$, NO$_2$ and
acids\cite{taatjes2017criegee,mauldin2012new}. The high energy and
short lifetime of these species complicates the direct experimental
characterization of CIs.\\

\noindent
The smallest CI is formaldehyde oxide (H$_2$COO).\cite{lee:2015} Key
to more detailed laboratory studies of H$_2$COO was its successful
{\it in situ} generation using photolysis of CH$_2$I$_2$ in
O$_2$.\cite{welz:2012} Computational work some 10 years ago
proposed\cite{samanta2014quantum,dawes2015uv,su2013infrared} that
energized H$_2$COO can decompose to HCO+OH and H$_2$CO+O$(^3 {\rm P})$
or isomerize to dioxirane. Dissociation to the HCO+OH channel proceeds
either via hydrogen-transfer (HT) to form the linear HCOOH isomer
(high-energy route, TS2),\cite{MM.h2coo:2024} or through TS1 and
intermediates such as dioxirane and formic acid following the
low-energy pathway, see Figure \ref{fig:diagram}. Further
intermediates along the low-energy pathway are cyc-H$_2$CO$_2$ and the
di-radical OCH$_2$O. Additional decomposition products include
CO$_2$+H$_2$ and H$_2$O+CO, see Figure \ref{fig:diagram}.\\

\noindent
Excitation of internal vibrational modes for decomposition reactions
(``vibrationally induced photodissociation'') in the atmosphere has
been proposed for a number of compounds, including HONO, HONO$_2$, or
HO$_2$NO$_2$.\cite{Donaldson:1997} These species absorb visible
radiation and induce vibrational overtone transitions to states with
several quanta in the OH stretching vibration. For H$_2$SO$_4$
(sulfuric acid) vibrationally induced reactivity by exciting the
OH-stretch mode was implicated in photodissociation dynamics forming
SO$_3$ and water from experiments\cite{Vaida:2003} and molecular
dynamics (MD)
simulations.\cite{Miller:2006,yosa:2011,reyes.pccp.2014.msarmd}
Nevertheless, although cavity ring-down spectroscopy successfully
probed the asymmetric OH stretching vibration of H$_2$SO$_4$ with
$\nu_9 = 4$ and $\nu_9 = 5$, vibrationally induced photodissociation
dynamics has as yet not been observed directly.\cite{Feierabend:2006}
Similarly, HFSO$_3$ and HClSO$_3$ have been characterized
experimentally\cite{lane:2007,vaida:2007} and
computationally\cite{MM.hso3cl:2016,MM.hso3f:2017} as potential
candidates for OH-stretch induced photodissociation.\\

\noindent
Vibrationally induced reactivity was also employed to initiate a
sequence of chemical transformations in the next-larger CI, {\it
  syn-}CH$_3$CHOO.\cite{fang:2016,fang:2016deep} Rates for appearance
of OH were measured from unimolecular dissociation by vibrationally
activating \textit{syn}-CH$_3$CHOO with energies equivalent to
approximately two quanta in the CH-stretch vibration. This energy is
close to the barrier for HT towards vinyl-hydroxyperoxide which
eventually can lead to OH elimination. Computationally, the entire
reaction pathway of energized {\it syn-}CH$_3$CHOO to OH(X$^2 \Pi$)
elimination was followed using a neural network-represented potential
energy surface (PES)\cite{MM.physnet:2019} based on complete active
space second order perturbation theory (CASPT2) calculations using the
cc-pVDZ basis set.\cite{MM.criegee:2021,MM.criegee:2023} In addition
to OH-elimination, OH-roaming and formation of glycolaldehyde was
found as an alternative reaction pathway.\cite{MM.criegee:2023} By way
of analyzing the translational kinetic energy distribution it was also
demonstrated\cite{MM.criegee:2023} that the entire reaction pathway
from reactant to product via non-equilibrium vibrational excitation
needs to be followed to obtain a quantitative understanding of the
measurements.\cite{lester:2016}\\

\noindent
In the present work, the end-to-end reaction dynamics of the smallest
CI, H$_2$COO, following vibrational excitation of internal vibrational
modes is considered to characterize formation of molecular products
along the low-energy pathway. Recently, the first step of the
vibrationally induced photodissociation dynamics of H$_2$COO was
investigated computationally for the high- and low-energy
pathways.\cite{MM.h2coo:2024} This work reported a machine
learning-based PES (ML-PES) based on CASPT2 reference data that covers
both products: dioxirane (through TS1) and linear HCOOH (through
TS2). Following earlier
work\cite{lester:2016,MM.criegee:2021,MM.criegee:2023} on {\it
  syn-}CH$_3$CHOO, an initial non-equilibrium ensemble of H$_2$COO was
prepared through vibrational excitation of the CH-stretch mode
combined with a torsional motion to accelerate barrier crossing
(TS1).\cite{MM.h2coo:2024} This study concluded that both pathways are
accessible through vibrational excitation but that the low-energy
pathway is likely to be more effective for OH-formation. Following the
computational work, experiments were published that used excitation
with $\sim 2$ quanta in the CH-stretch vibration, with estimated
energies $\sim 1$ kcal/mol below the computed barrier from a composite
method.\cite{lester:2024} The energy-dependent rate $k(E)$ for
OH-formation along the low-energy pathway from equilibrium statistical
unimolecular reaction theory (RRKM) overestimated the measured rate at
an excitation energy of 6266 cm$^{-1}$ by two orders of
magnitude. Furthermore, it was concluded that the first step
(formation of cyc-H$_2$CO$_2$) must proceed via tunneling exclusively
because the experimental excitation energy (17 - 18 kcal/mol) is below
the computed barrier (19.46 kcal/mol) at the HEAT-345(Q)$_\Lambda$
level of theory. However, this study focused entirely on OH-formation
and other possible and low-energy molecular reaction product states
were not monitored. In contrast, the present work quantitatively
reports on all three accessible asymptotes that yield molecular
fragments: HCO+OH, CO$_2$+H$_2$, and H$_2$O+CO.\\

\noindent
First, the methods used are introduced, followed by a validation of
the ML-PES. Then, product state distributions are reported and
geometrical features of the reactive trajectories are
analyzed. Finally, the results are discussed in a broader context and
conclusions are drawn.\\

\section{Methods}

\subsection{Construction of the Reactive ML-PES}
\begin{figure} [H]
    \centering \includegraphics[width=0.8\linewidth]{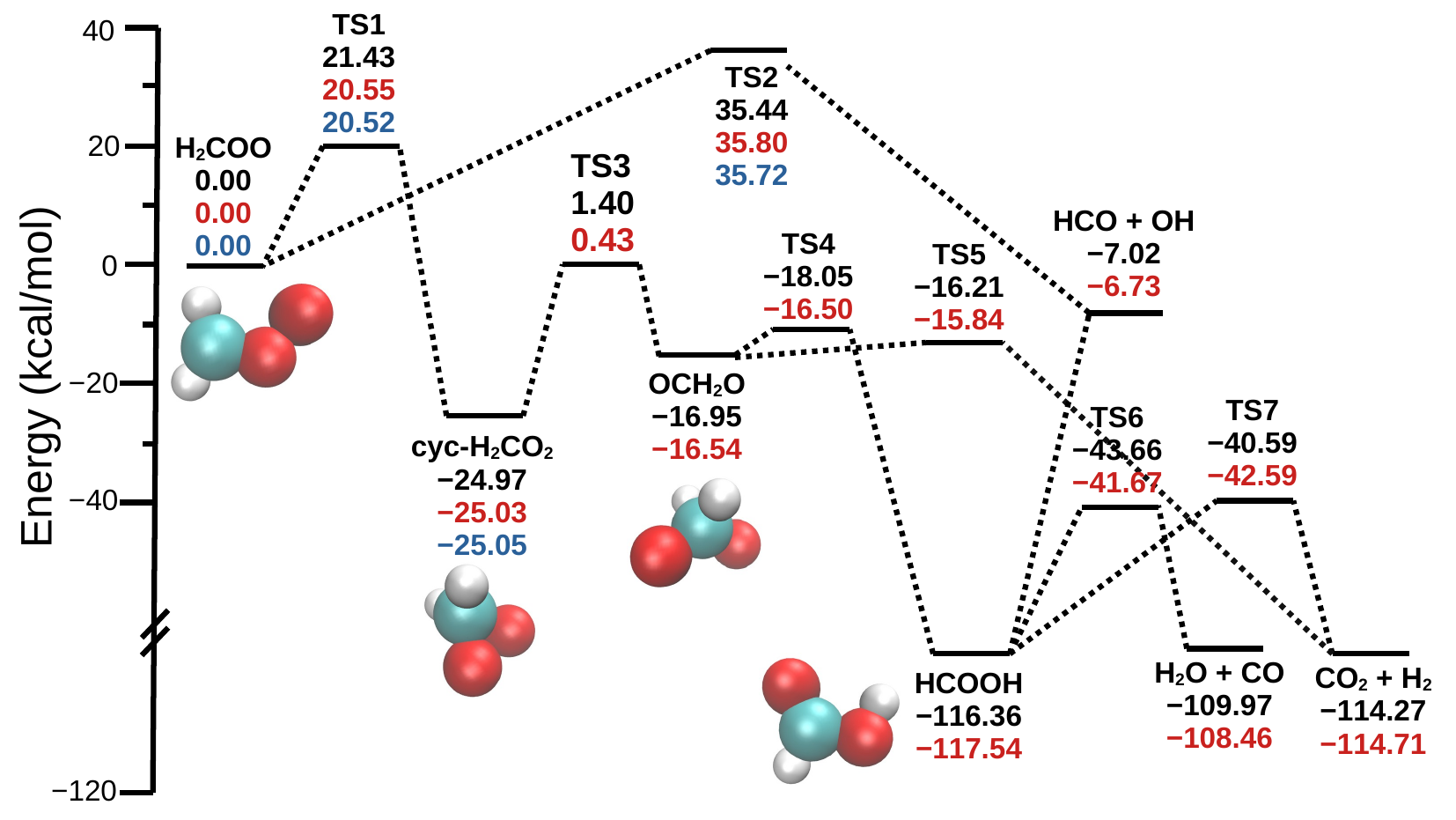}
    \caption{Schematic PES for the unimolecular decay of
      H$_2$COO. Black, red and blue numbers are for CASPT2/aVTZ
      calculations, PES2025 from the present work (PES2025), and
      PES2024,\cite{MM.h2coo:2024} respectively, all without zero
      point contributions. All molecular decomposition products
      CO$_2$+H$_2$, H$_2$O+CO, and HCO+OH were observed in the present
      simulations. The first barrier (TS1) for the reaction leading to
      CO$_2$+H$_2$ or H$_2$O+CO is 20.55 kcal/mol above H$_2$COO. This
      compares with reported barrier heights (without ZPE corrections)
      from other high-level calculations\cite{lester:2024} of 20.11,
      20.20 and 20.19 kcal/mol at the level of MRCI+Q/cc-pVDZ,
      CCSDT(Q)$_\Lambda$/cc-pVDZ and CCSDTQ/cc-pVDZ. The ZPE-corrected
      TS1 from PES2025 is 19.76 kcal/mol compared with 19.46 kcal/mol
      using the HEAT-345(Q)$_\Lambda$ method.\cite{lester:2024}}
    \label{fig:diagram}
\end{figure}

In the present work, the ML-PES covering the entire reaction pathway
between H$_2$COO and molecular products including CO$_2$+H$_2$,
H$_2$O+CO, and HCO+OH radicals was constructed, see Figure
\ref{fig:diagram}. Starting from an earlier data
set\cite{MM.h2coo:2024} which considered only the transition between
H$_2$COO and cyc-H$_2$CO$_2$ (5162 geometries with energies and
forces), energies and forces for additional states/geometries
including the di-radical OCH$_2$O intermediate, HCOOH (formic acid),
the CO$_2$+H$_2$, H$_2$O+CO, and HCO+OH products and all transition
states (TSs) connecting these states were determined (Figure
\ref{fig:diagram}). All necessary electronic structure calculations
were carried out at the CASPT2/aug-cc-pVTZ (CASPT2/aVTZ) level of
theory using the MOLPRO suite of codes.\cite{molpro:2020} The active
space for the complete active space SCF (CASSCF) calculations
consisted of 12 electrons in 11 orbitals.\\

\noindent
For extending the data set (training, validation and test), additional
geometries were generated from a combination of MD simulations at the
GFN2-xTB level\cite{bannwarth2019gfn2} as available in the Atomic
Simulation Environment (ASE),\cite{larsen2017atomic} normal mode
sampling around local minima, adaptive sampling using independently
trained models, and random displacement with a maximum amplitude of
0.3 \AA\/ around the Intrinsic Reaction Coordinate (IRC) geometries
for the decomposition paths. This resulted in 9201 new
geometries. Along with the original data set\cite{MM.h2coo:2024}
containing 5162 reference structures, the new data set comprised 14363
structures covering the entire reaction path, as shown in Figure
\ref{fig:diagram}, for which energies and forces were calculated at
the CASPT2/aVTZ level.\\

\noindent
PhysNet\cite{MM.physnet:2019} models were trained using an 80/10/10
split of the entire data set into training, validation, and test sets,
which was used consistently throughout. The hyperparameters for all
models trained in the present work were as follows. The NN consisted
of 5 modules, each atom $i$ was represented as a feature vector of
length 128, and the atomic feature vector for atom $i$ was iteratively
updated using information from neighbouring atoms $j$ within $r_{\rm
  cut} = 10$ \AA\/ using message-passing. The loss function
\begin{equation}
\begin{split}
L & = w_E|E-E^{\rm
  ref}|+\frac{w_F}{3N}\sum_{i=1}^N\sum_{\alpha=1}^3|-\frac{\partial
  E}{\partial r_{i,\alpha}}-F_{i,\alpha}^{\rm ref}|\\ &
+w_Q|\sum_{i=1}^Nq_i-Q^{\rm
  ref}|+\frac{w_p}3\sum_{\alpha=1}^3|\sum_{i=1}^Nq_ir_{i,\alpha}-p_\alpha^{\rm
  ref}|+L_{\rm nh}
\end{split}
\end{equation}
was minimized using AMSGrad.\cite{reddi2019convergence} Here, $E^{\rm
  ref}$ and $Q^{\rm ref}$ are the quantum chemical reference energy
and total charge, $p_\alpha^{\rm ref}$ represents the Cartesian
components of the reference dipole moment, $F_{i,\alpha}^{\rm ref}$
acting on atom {\it i}, and $r_{i,\alpha}$ is the $\alpha$th Cartesian
coordinate of atom {\it i}. The hyperparameters were $w_E = 1$
eV$^{-1}$ $w_F$ = 52.9177 \AA/eV, $w_Q$ = 14.3996 e$^{-1}$, and
$w_p$=27.2113 D$^{-1}$.\\

\noindent
Analysis of the first trained models indicated the presence of 486
structures for which the difference between reference calculations and
model predictions exceeded 10 kcal/mol and were classified as
``outliers''. These data points were removed and for the final data
set, containing 13877 geometries, four independent NN-PESs were
trained. Outliers occurred for all relevant states, including
cyc-H$_2$CO$_2$, di-radical OCH$_2$O intermediates, HCOOH,
CO$_2$+H$_2$, H$_2$O+CO, and HCO+OH. A particularly challenging
species was the dissociation into HCO+OH, for which an appreciable
fraction of the CASPT2 calculations did not converge. However, there
was still a sufficient number of structures $(\sim 600)$ for which
converged CASPT2 reference data was obtained, allowing to describe the
asymptotic region in a meaningful fashion, see Figure
\ref{sifig:hco_oh_cas}.\\

\subsection{MD Simulations and Analysis}
All production MD simulations were performed using the CHARMM and
pyCHARMM
programs\cite{brooks2009charmm,buckner2023pycharmm,MM.charmm:2024}
together with interfaces to
PhysNet.\cite{MM.pycharmm:2023,MM.asparagus:2025} Initial conditions
for the non-equilibrium simulations (see below) were generated from an
equilibrium 1 ns $NVE$ simulation after heating H$_2$COO to 300 K for
200 ps and equilibrating for 50 ps.\cite{MM.h2coo:2024} The time step
in the MD simulations was $\Delta t = 0.1$ fs to conserve total energy
as the bonds involving hydrogen atoms were flexible. From 10
independent simulations the coordinates and velocities were extracted
every 100 fs which yields a total of $10^6$ starting structures for
the non-equilibrium simulations described next.\\

\noindent
The non-ZPE-corrected barrier height between H$_2$COO and
cyc-H$_2$CO$_2$ is $\sim 20.6$ kcal/mol, see Figure
\ref{fig:diagram}. Initially, and following previous
work,\cite{MM.h2coo:2024} the dynamics was initiated through
vibrational excitation along the
3$\nu_{\mathrm{CH}}$+1$\nu_{\mathrm{COO}}$ combination mode (3 quanta
in the CH-stretch and 1 quantum in the COO bend) with a total energy
content of 25.5 kcal/mol. This was found to lead to transition towards
cyc-H$_2$CO$_2$ on the 10 to 1000 ps time scale. As will be discussed
further below, excitation closer to the TS1-barrier also leads to the
three final products (CO$_2$+H$_2$, H$_2$O+CO, and HCO+OH) but on
considerably longer time scales. Unless otherwise indicated, the
maximum simulation time was 1 ns and runs for which dissociated
products were reached earlier were
stopped. Experimentally,\cite{lester:2024} the overtone CH stretch
region around 2$\nu_{\mathrm{CH}}$ with an energy content of 17 to 18
kcal/mol was used to start the non-equilibrium dynamics.\\

\noindent
In order to associate individual structures from the MD simulations to
particular states along the reaction pathway (see Figure
\ref{fig:diagram}), specific geometrical criteria were defined for
approximate assignment. These criteria are summarized in Table
\ref{sitab:criteria}. As an example for such an assignment, if both
CH-separations are shorter than 1.5 \AA\/ candidate structures are
H$_2$COO, cyc-H$_2$CO$_2$ or OCH$_2$O. Next, if both $\theta_{\rm COO}
< 90^\circ$ H$_2$COO is excluded. Finally, the OCO angle determines
whether the structure is assigned to cyc-H$_2$CO$_2$ or OCH$_2$O.\\

\section{Results}

\subsection{Validation of the ML-PES}
First, the quality of the trained ML-PESs is discussed. For this, the
mean absolute error (MAE) and the root mean squared error (RMSE) for
energies and forces between the reference CASPT2 calculations and the
trained PhysNet models are considered, see Table
\ref{tab:evaluation_1387}.\\

\begin{table} [H]
    \centering
    \begin{tabular}{c|c|c|c|c}
    \hline Evaluation & ${\rm MAE}(E)$ & ${\rm RMSE}(E)$ & ${\rm
      MAE}(F)$ & ${\rm RMSE}(F)$ \\ \hline Model1 & 1.33 & 2.00 & 1.01
    & 3.08\\ Model2 & 1.43 & 2.19 & 1.09 & 4.49\\ Model3 & 1.40 & 2.10
    & 1.10 & 3.27\\ Model4 & 1.40 & 2.08 & 1.14 &
    3.67\\ PES2024\cite{MM.h2coo:2024} & 0.41 & 0.85 & 0.75 &
    3.54\\ \hline
    \end{tabular}
    \caption{Test set errors ($\sim 1400$ structures) of the trained
      PhysNet models. Energies in kcal/mol and forces in
      kcal/mol/\AA. The performance of PES2024 is also shown. The
      present data set covers an energy range of 280 kcal/mol compared
      with 160 kcal/mol from PES2024.\cite{MM.h2coo:2024}}
    \label{tab:evaluation_1387}
\end{table}

\begin{figure} [H]
    \centering \includegraphics[width=0.9\linewidth]{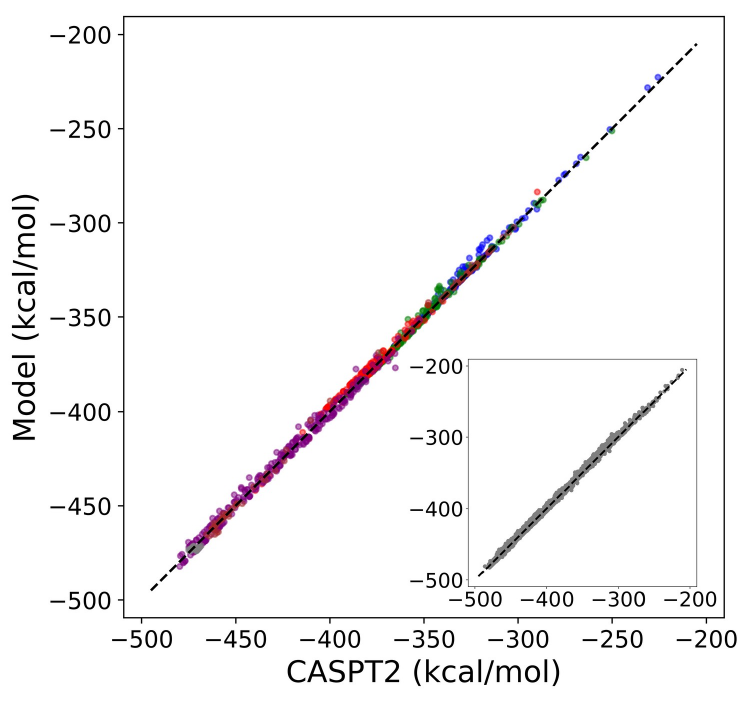}
    \caption{Correlation between the CASPT2/aVTZ energies and PES2025
      energies. Main view is for the test data set with $\sim 1400$
      structures. Blue: H$_2$COO, Orange: HCO+OH, Green:
      cyc-H$_2$CO$_2$, Red: di-radical OCH$_2$O, Purple: HCOOH, Brown:
      CO$_2$+H$_2$, Gray: H$_2$O+CO. Inset shows the training data set
      with $\sim 11100$ structures. $R^2$ = 0.9982/0.9984 for
      train/test data set. For the train data set, ${\rm MAE}(E) =
      1.37$ kcal/mol, ${\rm RMSE}(E) = 2.06$ kcal/mol, ${\rm MAE}(F) =
      0.51$ kcal/mol/\AA\/, ${\rm RMSE}(F) = 2.71$ kcal/mol/\AA). For
      the test data set, see Table \ref{tab:evaluation_1387}. Outliers
      occur for all relevant states. The dashed black line is for
      perfect 1:1 correlation.}
    \label{fig:PES2025}
\end{figure}

\noindent
The four independent PhysNet models feature comparable MAE and RMSE
for both energies and forces. For the best-performing model, Model1,
${\rm MAE}(E) = 1.33$ kcal/mol and ${\rm RMSE}(E) = 2.00$ kcal/mol,
see Figure \ref{fig:PES2025}. The MAE on forces is ${\rm MAE}(F) =
1.01$ kcal/mol/\AA\/ and ${\rm RMSE}(F) = 3.08$ kcal/mol/\AA\/. This
compares with (${\rm MAE}(E) = 0.41$ kcal/mol, ${\rm RMSE}(E) = 0.85$
kcal/mol, ${\rm MAE}(F) = 0.75$ kcal/mol/\AA\/, ${\rm RMSE}(F) = 3.54$
kcal/mol/\AA) for PES2024.\cite{MM.h2coo:2024}\\

\noindent
When comparing the performance of Model1 - referred to as PES2025 in
the following - from the present work and PES2024 it should be noted
that they were conceived in somewhat different ways and cover a
different number of states and energy ranges. PES2024 was based on
CCSD(T) reference data and transfer learned to the CASPT2 level of
theory whereas PES2025 was directly trained on the CASPT2 reference
data. PES2025 covers a considerably larger number of states (7 states
and 6 TSs vs. 2 states and 1 TS) and a much wider energy range (280
kcal/mol vs. 160 kcal/mol) than PES2024.\cite{MM.h2coo:2024} A direct
comparison between PES2024 and PES2025 for the IRC crossing TS1 is
reported in Figure \ref{sifig:irc1}. Furthermore, recent work found
that training reference data from multi-reference calculations (MRCI
and CASPT2) can incur statistical errors that are larger by 1 or 2
orders of magnitude compared with models based on single-reference
data.\cite{MM.noise:2024} Hence, the performance of PES2025 is
consistent with that of PES2024 and given its much wider range of
validity the quality is remarkable. Figure \ref{sifig:2pes} provides a
direct comparison of PES2024 and PES2025 on the 5162 reference data
used for training PES2024.\\

\noindent
For the remainder of the present work, only PES2025 (Model1) was used
since it features the lowest MAE and RMSE. The energies of stationary
points from PES2025 and CASPT2 calculation are reported in Figure
\ref{fig:diagram}. Comparison between PES2025 and CASPT2 indicates
that the representation errors are within 1 kcal/mol for most
stationary points, except for HCOOH and H$_2$O+CO, for which the
errors are between 1 and 2 kcal/mol, and TS6 and TS7, for which the
errors are $\sim 2$ kcal/mol.\\

\noindent
The tight clustering of points along the diagonal $y=x$ black dashed
line in Figure \ref{fig:PES2025} suggests high predictive
accuracy. This is also reflected in high correlation coefficients $R^2
= 0.9982/0.9984$ for train/test data set which further support strong
model performance and indicates that the model captures the underlying
trends in the data accurately. The main view displays the test data
set, consisting of approximately 1400 structures, with different
colors representing each species. Performance on the training data set
($\sim 11,100$ structures) is shown in the inset.\\

\begin{figure} [H]
    \centering \includegraphics[width=0.9\linewidth]{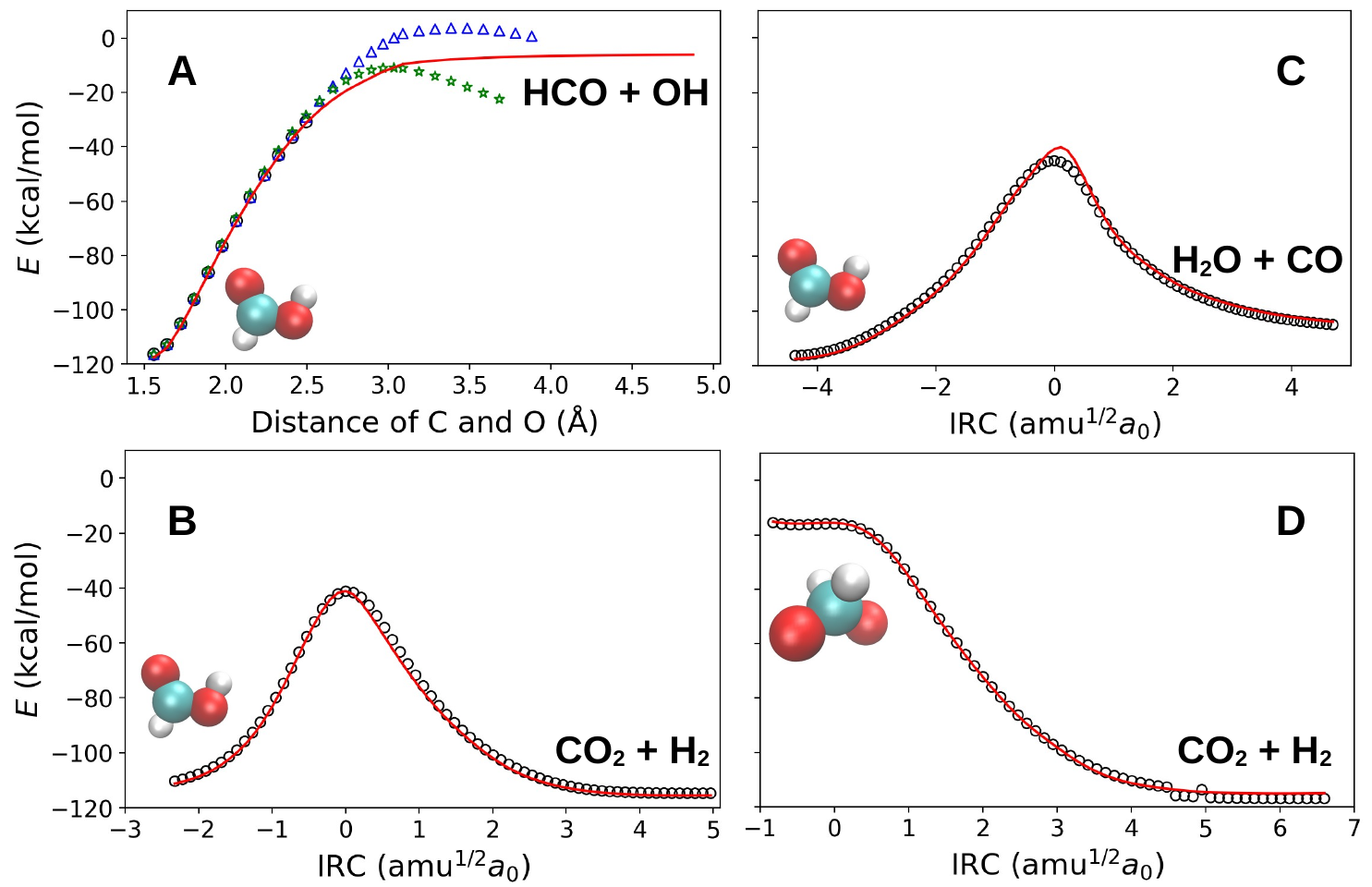}
    \caption{Energy profiles from using PES2025 (red line) for the four
      decomposition pathways. Panel A: The HCO+OH dissociation pathway
      starting from formic acid. The reaction coordinate is the
      C$_{\rm HCO}$--O$_{\rm OH}$ separation $d_{\rm CO}$. The
      CCSD(T)/aVTZ (green stars) and MP2/aVTZ energies (blue circles)
      closely follow PES2025 up to $d_{\rm CO} = 2.5$ \AA\/ after which
      multi reference effects lead to unphysical behaviour. Panels B
      and D: formation of CO$_2$+H$_2$ from formic acid (barrier
      height 43 kcal/mol) and OCH$_2$O (release of 98 kcal/mol),
      respectively. Panel C: formation of H$_2$O+CO from formic
      acid. Open circles in panels B to D are from IRC calculations at
      the CASPT2/aVTZ level of theory.}
    \label{fig:irc}
\end{figure}

\noindent
The photodissociation dynamics between vibrationally excited H$_2$COO
and molecular products included formation of CO$_2$+H$_2$, H$_2$O+CO,
and HCO+OH. The IRCs between constitutional isomers (structural
isomers) of H$_2$COO and corresponding molecular fragments are
reported in Figures \ref{fig:irc}A to C together with the reaction
path between OCH$_2$O and CO$_2$+H$_2$ (panel D). For dissociation
into HCO+OH (Figure \ref{fig:irc}A) the energy profile along the
C$_{\rm HCO}$--O$_{\rm OH}$ separation agrees well between PES2025
(red line) and the CASPT2 (black cross), CCSD(T) (green star), and MP2
(blue circles) calculation up to a separation of 2.5 \AA\/. However,
beyond 2.5 \AA\/, the CASPT2 calculations become more difficult to
converge due to the electronic structures, and there is no guarantee
that the chosen active space remains suitable for all geometries. In
contrast, CCSD(T) and MP2 calculations do converge but produce
unreliable results, as a single-reference treatment is inadequate in
this region. Nevertheless, it was possible to converge energies for a
sufficiently large number of perturbed geometries along the IRC to
train PhysNet, see Figure \ref{sifig:hco_oh_cas}. Evidently, PES2025
(red line in Figure \ref{fig:irc}A) yields a meaningful energy profile
to form HCO+OH.\\

\noindent
Figure \ref{fig:irc}B reports dissociation to CO$_2$+H$_2$ coming from
formic acid. The barrier for this process is 43 kcal/mol above formic
acid and PES2025 very closely matches the IRC at the CASPT2 level
(open circles). The same is observed for dissociation to H$_2$O+CO
which is shown in Figure \ref{fig:irc}C. Here, the barrier height is
42 kcal/mol above HCOOH and the profile around the top of the barrier
towards molecular products is still meaningful but not as good as for
dissociation into CO$_2$+H$_2$. Finally, Figure \ref{fig:irc}D shows
the IRC between OCH$_2$O and CO$_2$+H$_2$. This process is downhill
and the solid line (PES2025) accurately describes the reference
data. Notably, for large values of the reaction coordinate, PES2025 is
smooth whereas the CASPT2 calculations for large CO$_2$--H$_2$
separations show some jitter due to slight numerical instabilities. In
conclusion, for all 4 pathways in Figure \ref{fig:irc} the energy
profiles from using the ML-PES closely follow those from CASPT2/aVTZ
calculations.\\

\begin{figure} [H]
    \centering \includegraphics[width=0.8\linewidth]{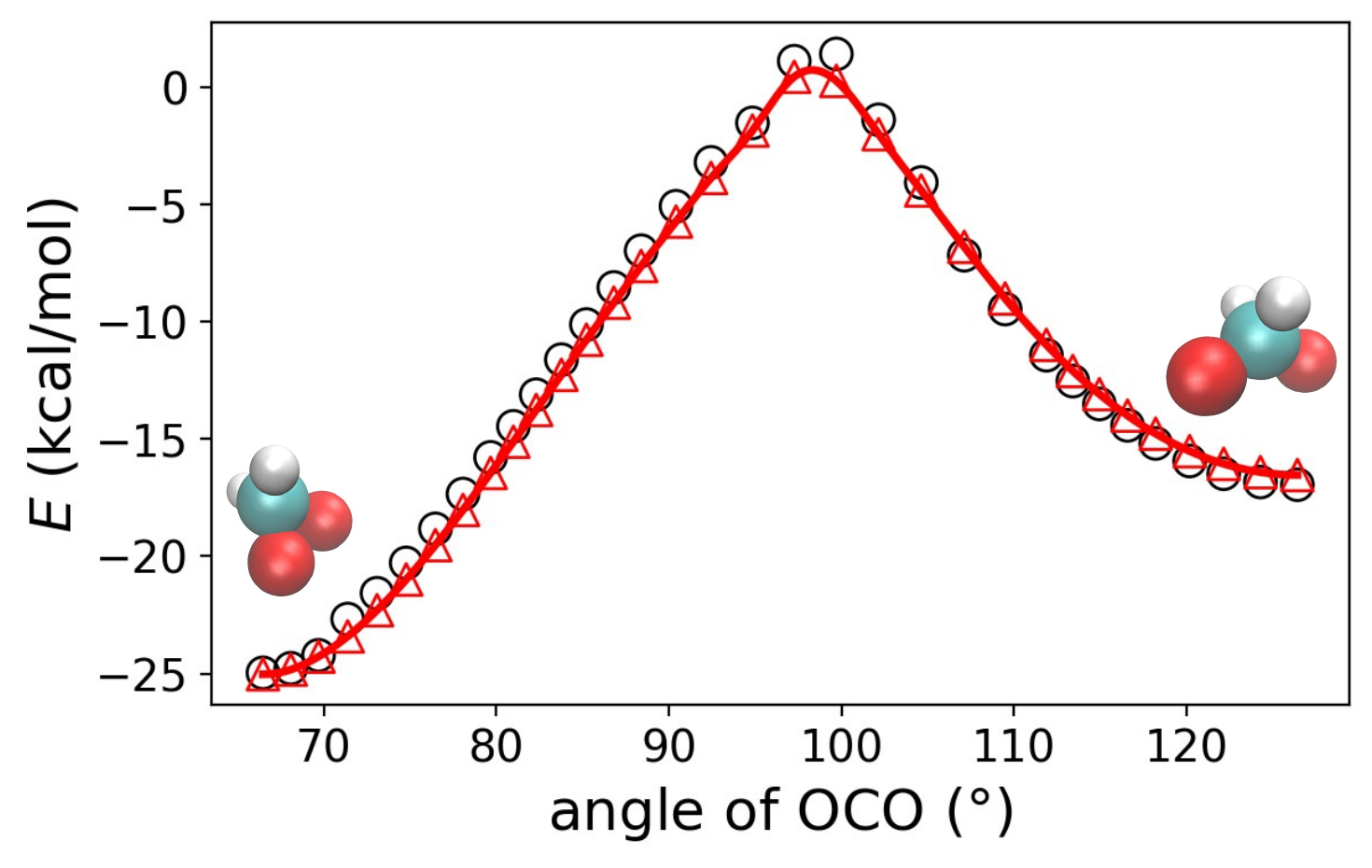}
    \caption{The energy profile between cyc-H$_2$CO$_2$ (left) and
      di-radical OCH$_2$O (right) from a NEB calculation in ASE using
      PES2025 as the energy function. Energies from PES2025 (red line
      and triangles) and CASPT2/aVTZ (black circles) are compared.
      The largest geometrical change between cyc-H$_2$CO$_2$ and
      di-radical OCH$_2$O concerns the OCO angle: $\angle {\rm OCO} =
      66^{\circ}$ for cyc-H$_2$CO$_2$, and $\angle {\rm OCO} =
      126^{\circ}$ for di-radical OCH$_2$O. At the TS ($\angle {\rm
        OCO} \sim 100^{\circ}$ the barrier heights are 25.5 / 26.4
      kcal/mol from PES2025 / CASPT2. The CH and CO bond lengths for
      the cyc-H$_2$CO$_2$/TS/OCH$_2$O structures are [1.08/1.11/1.12]
      \AA\/ and [1.39/1.37/1.32] \AA\/, and the HCH angles are
            [117/108/98]$^{\circ}$, respectively.}
    \label{fig:diox}
\end{figure}

\noindent
As a further validation of PES2025, the energy profile between
cyc-H$_2$CO$_2$ and the di-radical OCH$_2$O (see Figure
\ref{fig:diagram}) was evaluated. Despite intensive effort to obtain
this transition path directly from electronic structure calculations
at the CASPT2/aVTZ level, using IRC-following methods,\cite{irc:1990}
this was not feasible. As an alternative, a nudged elastic band (NEB)
calculation\cite{henkelman2000climbing} using PES2025 was carried out
in ASE from which 9 intermediate structures along with optimized
cyc-H$_2$CO$_2$ and di-radical OCH$_2$O were obtained. To increase the
density of the mesh, new structures were generated by averaging the
coordinates between two adjacent NEB-generated geometries. Energies
for all the structures determined from PES2025 (red triangle) and
CASPT2 (black circle) calculations are compared in Figure
\ref{fig:diox}. Along the NEB-path, the OCO angle between
cyc-H$_2$CO$_2$ and OCH$_2$O changes from $66^{\circ}$ to
$126^{\circ}$ with a barrier height of $\sim 26$ kcal/mol at $\angle
{\rm OCO} \sim 100^{\circ}$. This barrier height compares and is
consistent with earlier reported values of 21.1 kcal/mol and 19.2
kcal/mol at the UCCSD(T)/CBS//UCCSD/aug-cc-pVTZ\cite{nguyen:2015} and
DLPNO-CCSD(T)/CBS(aug-cc-pV5Z,aug-cc-pV6Z)//CASSCF(8,8)/def2-TZVP\cite{peltola2020time}
levels of theory.\\

\noindent
For a more global validation of the ML-PES, Diffusion Monte Carlo
(DMC) simulations using PES2025 were run to check the ML-PES for
holes.\cite{li:2021} Here, a ``hole'' was defined as a geometry for
which the energy predicted by PES2025 was lower than the global
minimum. Several independent DMC simulations using the unbiased
algorithm\cite{qu2021breaking,MM.fa:2022} were started from seven TSs
and four minima including H$_2$COO, cyc-H$_2$CO$_2$, di-radical
OCH$_2$O, and HCOOH, see Figure \ref{fig:diagram}. Each run used 3000
walkers and was propagated for a total of 55000 steps for each
simulation. For a total of $\sim 2.1 \times 10^9$ structures not a
single hole was detected which suggests that as per this test the
ML-PES is ``hole free''.\\

\subsection{Exploratory MD Simulations}
The extensive simulations described in the next section indicated that
vibrationally excited H$_2$COO can follow at least three different
reaction pathways. Before considering and analyzing a statistically
significant number of trajectories, three representative examples of
photodissociation events are discussed and analyzed in more
detail. This was also done to further assess the quality of the
trained ML-PES from comparing PES2025-energies of MD-sampled
structures with reference calculations at the CASPT2/aVTZ level of
theory. These simulations were initialized by non-equilibrium
excitation of the 3$\nu_{\mathrm{CH}}$+1$\nu_{\mathrm{COO}}$
combination band. It should be noted that a typical single point
energy together with one force calculation at the CASPT2/aVTZ level
for H$_2$COO takes $\sim 5$ hours, i.e. running meaningful and
informative {\it ab initio} MD simulations on the 1 ns time scale at
this level of theory is infeasible.\\

\begin{figure} [h!]
    \centering
    \includegraphics[width=1.0\linewidth]{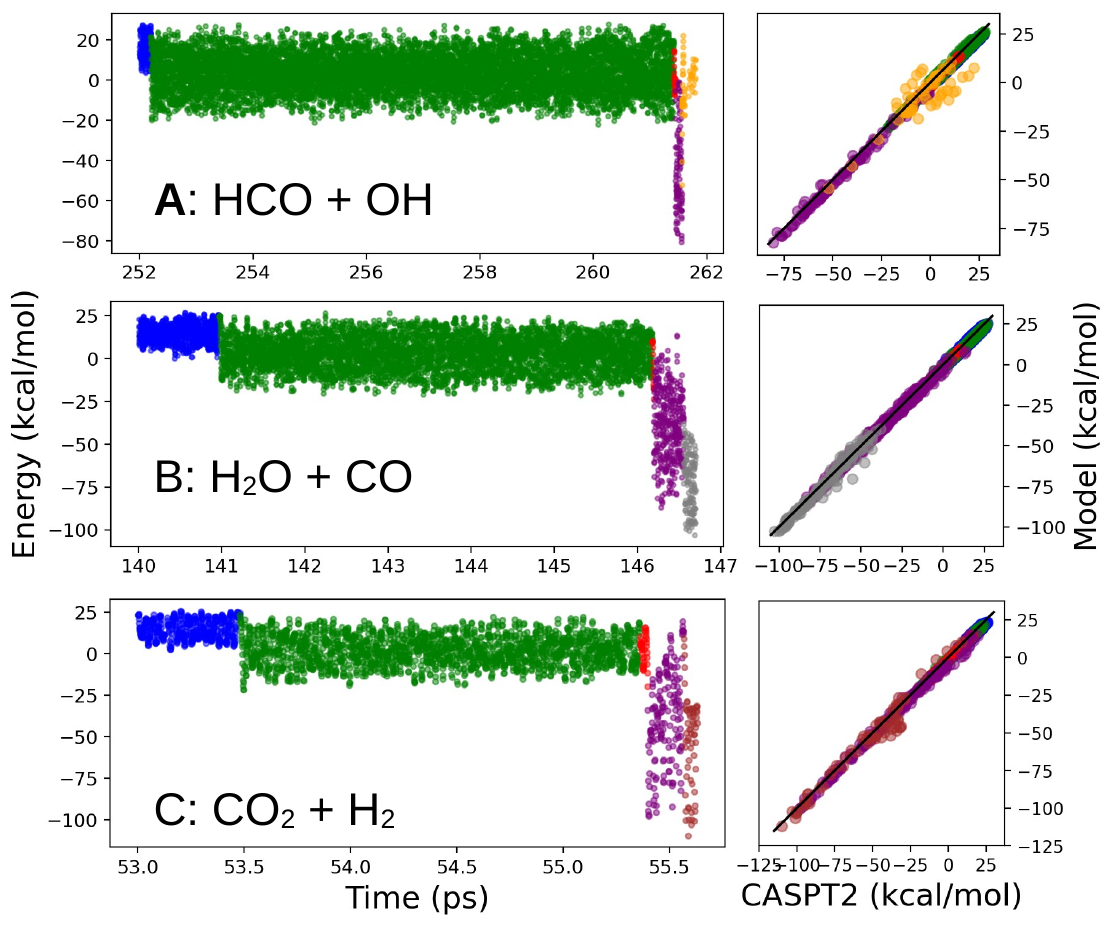}
    \caption{Typical trajectories for the three decomposition pathways
      to form HCO+OH (panel A), H$_2$O+CO (panel B), and CO$_2$+H$_2$
      (panel C). The color code for sampled geometries: Blue:
      H$_2$COO, Green: cyc-H$_2$CO$_2$, Red: di-radical OCH$_2$O,
      Purple: HCOOH, Orange: HCO+OH, Gray: H$_2$O+CO, and Brown:
      CO$_2$+H$_2$. Left column: PES2025 energies as a function of time
      for time frames during which relevant structures are sampled;
      right column: comparison of energies obtained from PES2025 and
      CASPT2. The values for $R^2$ and MAE/RMSE are reported in the
      paragraphs describing this figure.}
    \label{fig:three-traj}
\end{figure}

\noindent
The trajectory in Figure \ref{fig:three-traj}A leads to HCO+OH as the
reaction product. Starting from H$_2$COO, transitioning to
cyc-H$_2$CO$_2$ occurs at $\sim 252$ ps. After a lifetime of $\sim 9$
ps in this conformation, the di-radical OCH$_2$O appeared for 0.02 ps
followed by a transition to HCOOH which existed for $\sim 0.1$ ps.
Finally, decomposition to HCO+OH occurred. The correlation between the
PES2025 and CASPT2/aVTZ energies is shown in the right-hand panel of
Figure \ref{fig:three-traj}A: the MAE, RMSE and $R^2$, are 1.39
kcal/mol, 1.51 kcal/mol and 0.9963, respectively, which are consistent
with the performance of the ML-PES on the test data set, see Table
\ref{tab:evaluation_1387}.\\

\noindent
Decomposition to H$_2$O+CO is shown in Figure
\ref{fig:three-traj}B. At $\sim 141$ ps after non-equilibrium
excitation of H$_2$COO, cyc-H$_2$CO$_2$ is formed. The di-radical
OCH$_2$O appeared $\sim 5$ ps later and the transition to HCOOH
happened another 0.02 ps later. The lifetime of HCOOH was $\sim 0.4$
ps after which decomposition into H$_2$O+CO occurred. The MAE, RMSE
and $R^2$ between PES2025 and CASPT2/aVTZ energies are 1.50 kcal/mol,
1.62 kcal/mol and 0.9982, respectively, see right-hand panel.\\

\noindent
Finally, formation of reaction products CO$_2$+H$_2$ is considered in
Figure \ref{fig:three-traj}C. Within $\sim 53.5$ ps of vibrational
excitation of H$_2$COO, cyc-H$_2$CO$_2$ is formed with a lifetime of
$\sim 2$ ps to give the di-radical OCH$_2$O. This species lived for
less than 0.03 ps followed by HCOOH which exists for $\sim 0.2$ ps
after which decay to CO$_2$+H$_2$ occurs. For this trajectory, the
statistical measures between PES2025 and CASPT2/aVTZ energies are MAE,
RMSE and $R^2$ are 1.67 kcal/mol, 1.94 kcal/mol and 0.9962,
respectively.\\

\noindent
It should be noted that in all cases considered, the MAE and RMSE are
dominated by the reaction product states. Hence, if final internal
state distributions (vibrational, rotational) of the fragments is of
interest - which is outside the scope of the present work - further
validation of the total energy function, in particular for monomer
deformation energies will be required.\\

\subsection{MD Simulations Covering the Entire Pathway}
For a comprehensive characterization of the H$_2$COO unimolecular
decomposition from molecular simulations an ensemble of initial
structures was vibrationally excited in a non-equilibrium
fashion. Because the COO-bend coordinate has been found to be heavily
involved in the reaction coordinate leading from H$_2$COO to
TS1\cite{MM.h2coo:2024}, the $3 \nu_{\mathrm{CH}} + 1
\nu_{\mathrm{COO}}$ combination mode was used in the MD
simulations. This introduces an energy of 25.5 kcal/mol which is $\sim
5.0$ kcal/mol above the TS between H$_2$COO and cyc-H$_2$CO$_2$, see
Figure \ref{fig:diagram}. In this excitation scheme dwell times in the
H$_2$COO reactant state are on the 10 to 100 ps time scale. The
shortest reaction time is 20 ps which is, however, still considerably
longer than the time scale for partial relaxation of the CH-stretch
mode which is on the order of $\sim 5$ ps, see Figure \ref{sifig:ch}:
within 4 ps the amplitude of the CH-bond length decreases by 40 \%
which strongly diminishes the probability for crossing TS1 on short
(ps) time scales. On the other hand, the CH-stretch still samples a
considerably wider range of bond lengths compared with the equilibrium
distribution at 300 K (red trace).\\

\noindent
In order to characterize the temporal evolution of a vibrationally
excited ensemble of initial H$_2$COO conformations, 5000 independent
non-equilibrium trajectories were run. All 5000 trajectories were
followed for 1 ns or until diatomic fragments were formed. Out of the
5000 trajectories, 2186 remain in the reactant well for the entire 1
ns due to partial IVR at early times, see previous paragraph. The
remainder of the simulations progresses towards products on different
time scales. Of the reactive trajectories, 1616/966/77/29 produced
CO$_2$+H$_2$/H$_2$O+CO/HCO+OH/cyc-H$_2$CO$_2$. Finally, 126
trajectories led to HOCO+H (98 events) or CO$_2$+H+H (28
events). These final states differ significantly from the structures
that were used to train the ML-PES and were therefore excluded from a
detailed analysis.\\

\noindent
In terms of fractions of populations after 1 ns, 43.7 \% of the
trajectories remain in the H$_2$COO reactant state, formation of
CO$_2$+H$_2$ is the most populated product state (32.3 \%), followed
by H$_2$O+CO (19.3 \%), HCO+OH (1.6 \%), and cyc-H$_2$CO$_2$ (0.6
\%). A small fraction (2.5 \%) of the trajectories leads to other
reaction products, see above. Interestingly, the cyc-H$_2$CO$_2$
intermediate does not accumulate and is populated for $< 1$\% of the
trajectories. This makes cyc-H$_2$CO$_2$ only a ``stepping stone''
between reactant and products, despite the substantial barrier height
(25.5 kcal/mol) towards TS3 separating cyc-H$_2$CO$_2$ from di-radical
OCH$_2$O, see Figure \ref{fig:diox}. To estimate the statistical error
the unreacted fraction was considered. For a randomly selected subset
of 1000 out of the 5000 total trajectories the fraction that remained
in H$_2$COO was determined. This procedure was repeated 100 times
which yields an average fraction of $44.1 \pm 1.2$ \% which is close
to the ensemble average of 43.7 \%.\\

\begin{figure} [H]
    \centering \includegraphics[width=0.8\linewidth]{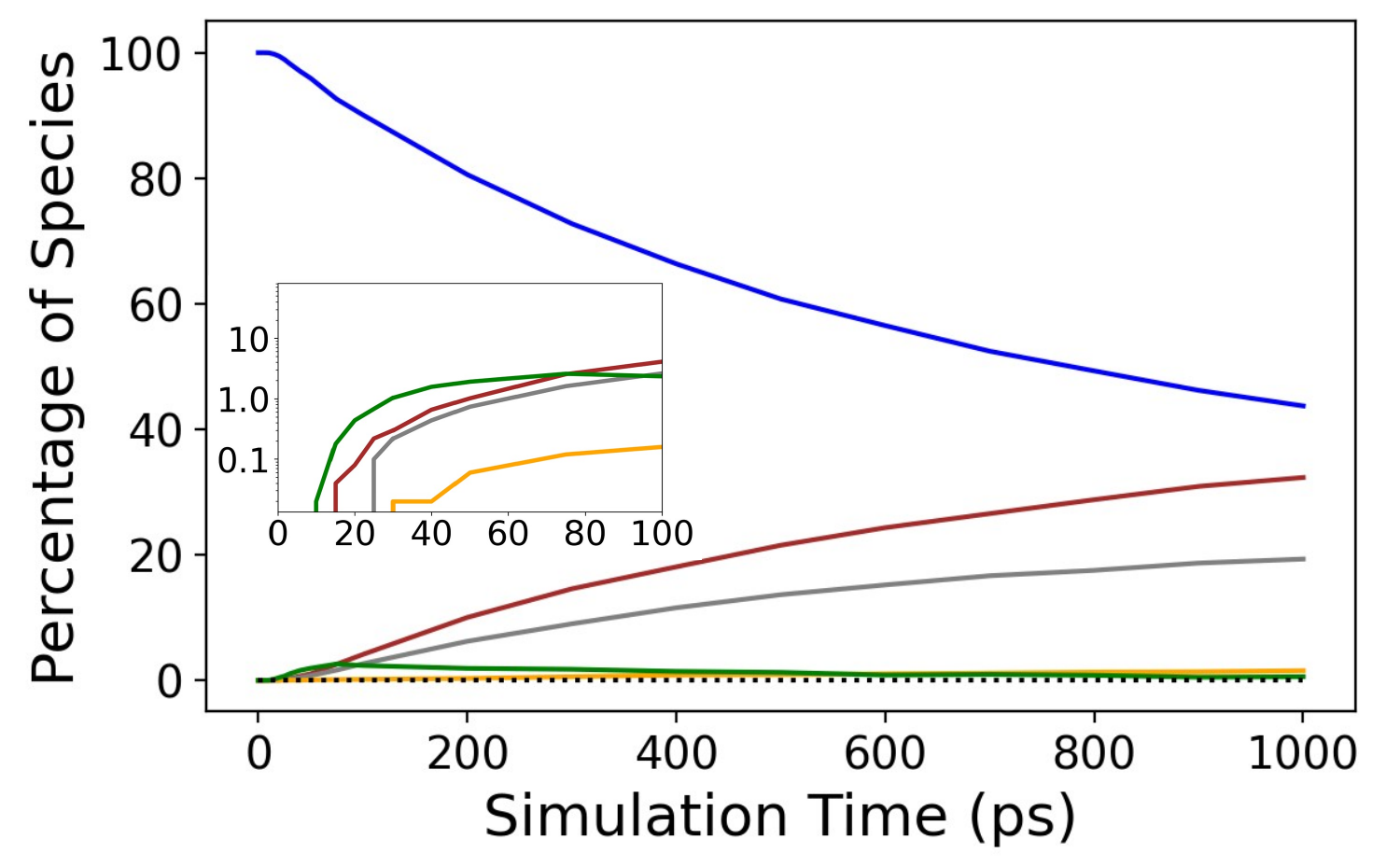}
    \caption{Changes in time-dependent populations for H$_2$COO
      (blue), cyc-H$_2$CO$_2$ (green), HCO+OH (Orange) CO$_2$+H$_2$
      (brown), and H$_2$O+CO (gray). Among the $\sim 2800$ reactive
      trajectories, the first
      cyc-$_2$CO$_2$/CO$_2$+H$_2$/HCOOH/H$_2$O+CO/HCO+OH appeared at
      9.8/11.8/16.1/21.0/26.6 ps. After 1 ns, the population of the
      products CO$_2$+H$_2$/H$_2$O+CO reach 32.3/19.3 \%, whereas 43.7
      \% remain in H$_2$COO and decay on considerably longer time
      scales. This was confirmed by continuing 100 unreacted
      trajectories for another 4 ns, for which only 12 trajectories
      remained as H$_2$COO, see Figure \ref{sifig:popu_5000}.}
    \label{fig:popu}
\end{figure}

\begin{figure} [H]
    \centering \includegraphics[width=0.8\linewidth]{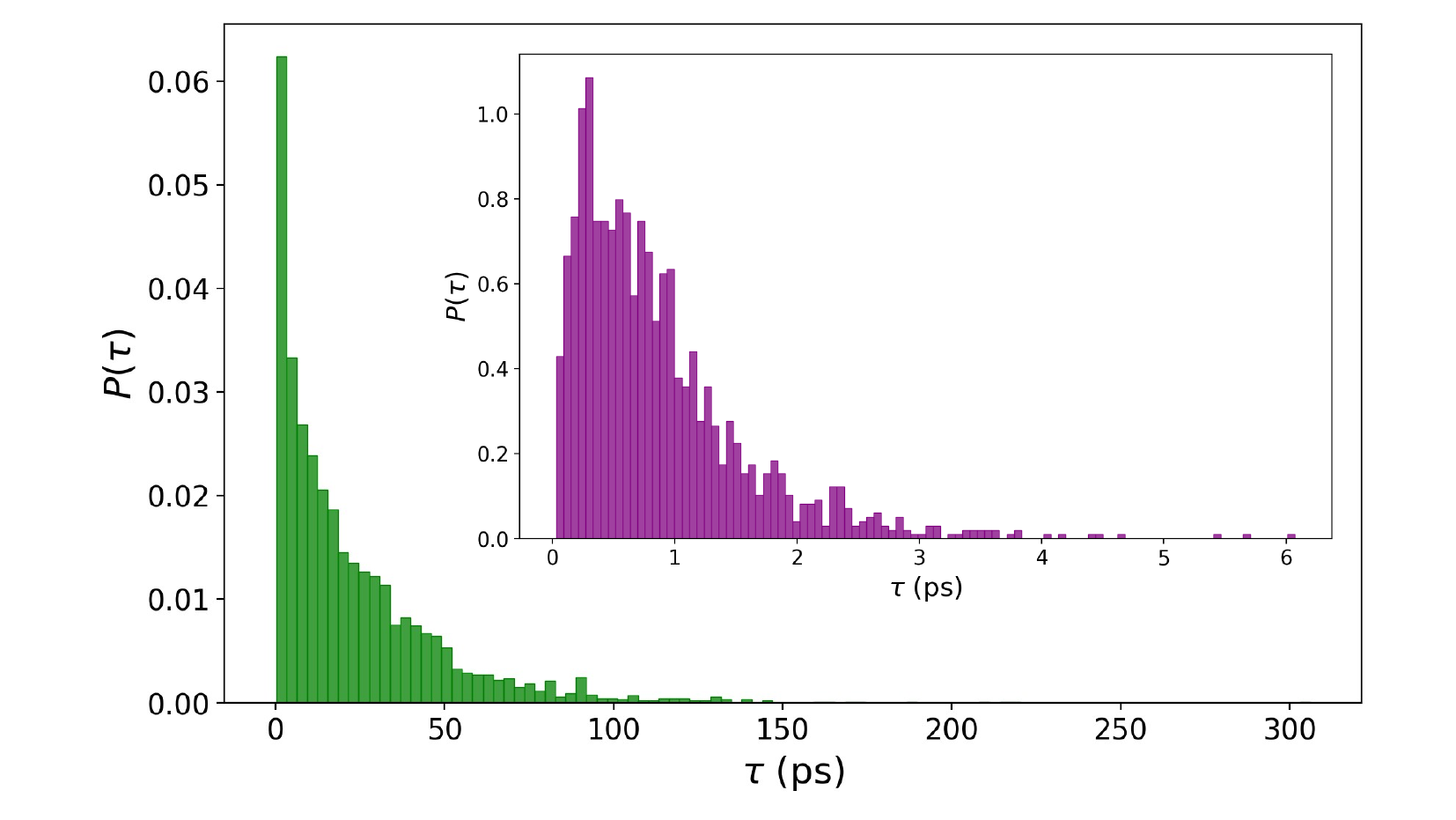}
    \caption{The normalized lifetime distributions of cyc-H$_2$CO$_2$
      from $\sim 2700$ reactive trajectories. Inset: Normalized
      lifetime distributions for HCOOH from $\sim 1600$ reactive
      trajectories.}
    \label{fig:life}
\end{figure}

\noindent
Going beyond final state branching ratios (after 1 ns), the
concentration (population) changes as a function of simulation time
for the different states and species of interest are shown in Figure
\ref{fig:popu}. As time progresses, reactant H$_2$COO (blue) gradually
decreases, whereas the final products CO$_2$+H$_2$ (brown), H$_2$O+CO
(gray), and HCO+OH (orange) gradually increase but on different time
scales. Formation and consumption of the cyc-H$_2$CO$_2$ intermediate
(green) is largely balanced and the species does not accumulate. Its
largest concentration reaches 2 \% (after 50 ps) but after a total
simulation time of 1 ns, the population has decreased to 0.6 \%. As
for the other two intermediates, di-radical OCH$_2$O and HCOOH, due to
their short lifetime (Figure \ref{fig:life}), their concentrations are
always $< 10^{-4}$ and not shown in Figure \ref{fig:popu}.\\

\noindent
Among the 5000 MD simulations, only 77 (1.6 \%) formed HCO+OH. This
explains the slow formation of OH products from CH$_2$OO observed in
the experiment with a reported time resolution of 4
ns.\cite{lester:2024} In addition, after 1 ns of simulation, only 0.6
\% cyc-H$_2$CO$_2$ are accumulated, although cyc-H$_2$CO$_2$ can live
up to 100 ps. The average lifetimes of cyc-H$_2$CO$_2$ and HCOOH are
23.2 and 0.87 ps, and their median lifetime is 14.9 and 0.69 ps,
respectively. The gradual decrease of [H$_2$COO] in Figure
\ref{fig:popu} indicates that the first step of the reaction, leading
from H$_2$COO to cyc-H$_2$CO$_2$, can occur at any time within the 1
ns simulation time. Regardless of when the concentration is
determined, the total concentrations of cyc-H$_2$CO$_2$ and HCOOH are
invariably $<3$ \%. As for the di-radical OCH$_2$O intermediate, due
to its unstable nature (low barriers TS4 and TS5), the lifetime is
typically $\sim 0.02$ ps.\\

\noindent
The lifetimes of some of the intermediates are broadly distributed,
see Figure \ref{fig:life}. The cyc-H$_2$CO$_2$ intermediate which is
only populated at the $<2$ \% level throughout the photodissociation
dynamics features lifetimes between 1 ps and 150 ps. On the other
hand, formic acid, which is the global minimum on the full reaction
landscape (see Figure \ref{fig:diagram}), lives only for a few ps on
average before it decays to products. This is rather surprising as the
stabilization energy towards products (TS6 and TS7) is still $\sim 80$
kcal/mol. The reason for such efficient product yield are at least
twofold: the high energy with which HCOOH is formed and the fact that
sufficient energy is contained along the relevant dissociative
coordinates. Examples for the temporal development of important and
intuitive 1-dimensional reaction coordinates for each of the 3 product
channels are given in Figure \ref{sifig:react-coor}.\\

\begin{figure} [H]
    \centering \includegraphics[width=0.8\linewidth]{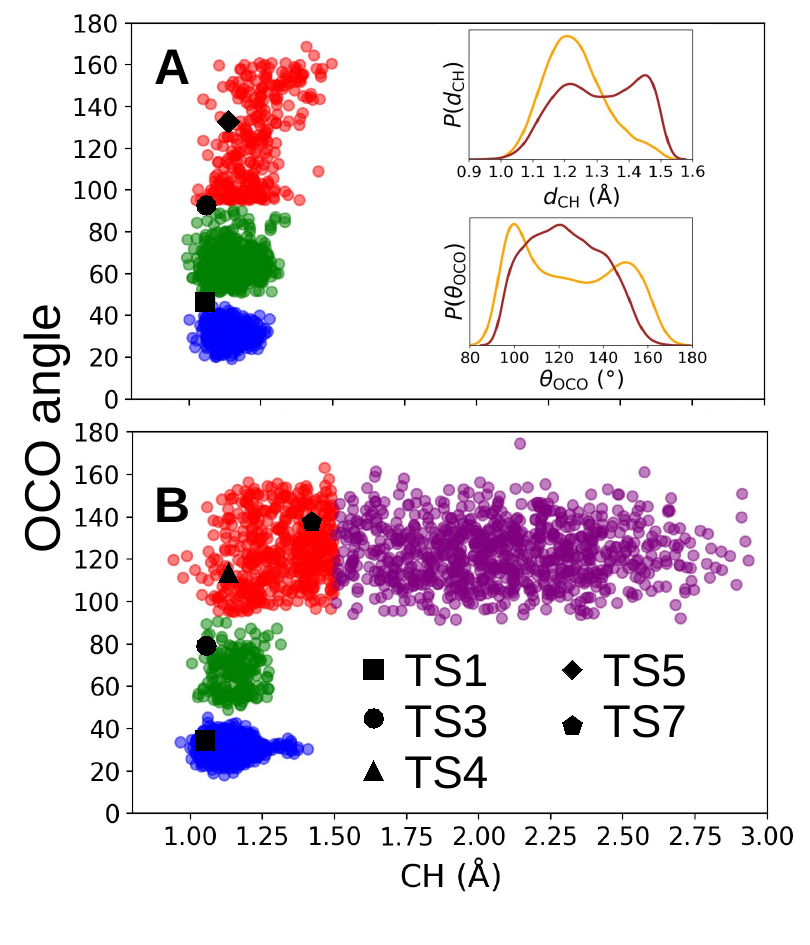}
    \caption{Bifurcating pathway connecting OCH$_2$O with
      CO$_2$+H$_2$. Panel A reports the distribution of OCO angle as a
      function of the CH distance from the reactive trajectories
      forming CO$_2$+H$_2$ directly from OCH$_2$O while panel B shows
      those passing through HCOOH. The color code: H$_2$COO (blue),
      cyc-H$_2$CO$_2$ (green), di-radical OCH$_2$O (red), and HCOOH
      (purple). The geometries of the three TSs in Figure
      \ref{fig:diagram} are marked, respectively. The insets show the
      CH distance and OCO angle distribution of samples of di-radical
      OCH$_2$O from the simulation trajectories. Orange lines are the
      geometries forming CO$_2$+H$_2$ directly from OCH$_2$O while
      Brown lines show the ones pass through HCOOH.}
    \label{fig:CH_OCO}
\end{figure}

\noindent
The present simulations also provide information about the reaction
path that is followed for individual reaction products. Formation of
CO$_2$+H$_2$ as the final product follows two different paths
(bifurcating pathway): out of the 1616 trajectories, 476 passed
through di-radical OCH$_2$O $\rightarrow$ TS4 $\rightarrow$ HCOOH
$\rightarrow$ TS7 $\rightarrow$CO$_2$+H$_2$ whereas 1140 follow the
route di-radical OCH$_2$O$\rightarrow$TS5$\rightarrow$CO$_2$+H$_2$,
sidestepping the deep well formed by formic acid. Rather, the shallow
transition region around TS5 which connects di-radical OCH$_2$O and
CO$_2$+H$_2$ is sampled. The geometrical space covered along these two
pathways is represented in Figure \ref{fig:CH_OCO}. For this, the
correlation between the C-H distance (the longer one of the two) and
the OCO angle for H$_2$COO (blue), cyc-H$_2$CO$_2$ (green), di-radical
OCH$_2$O (red), and HCOOH (purple) is analyzed. This provides a
low-dimensional projection of the configuration space characterizing
the trajectories for each of the reaction pathways.\\

\noindent
Figure \ref{fig:CH_OCO}A is for the pathway P$_{\rm TS5}$ passing
through TS5 whereas panel B reports the results for the pathway
P$_{\rm FA}$ through formic acid (HCOOH). The geometries of the TSs
are marked as black filled symbols. As is often the case for
vibrationally activated reactants,\cite{yosa:2011} the trajectories do
not necessarily pass through the exact geometries of the TSs,
especially for TS5 and TS7 that lead to the decomposition product
CO$_2$+H$_2$. For pathway P$_{\rm TS5}$ the CH-separation remains
centered around $\sim 1.2$ \AA\/ and the OCO angle increases as the
trajectories approach the product side. This differs considerably for
pathway P$_{\rm FA}$. Particularly for the di-radical OCH$_2$O
intermediate, one CH-bond needs to stretch to form HCOOH. This also
widens the red distribution associated with OCH$_2$O. The insets
compare the corresponding 1-dimensional distributions $P(d_{\rm CH})$
and $P(\theta_{\rm OCO})$ which also reflect these differences. The
sampling of the di-radical OCH$_2$O along P$_{\rm FA}$ feature longer
CH distances and a more pronounced population of $P(\theta_{\rm OCO})$
for $\theta_{\rm OCO} \sim 120^{\circ}$. This is consistent with the
fact that forming HCOOH from the di-radical OCH$_2$O involves an
H-atom ``roaming" process, and the OCO angle for the di-radical
OCH$_2$O and HCOOH are close to one another.\\

\begin{figure} [H]
    \centering \includegraphics[width=0.8\linewidth]{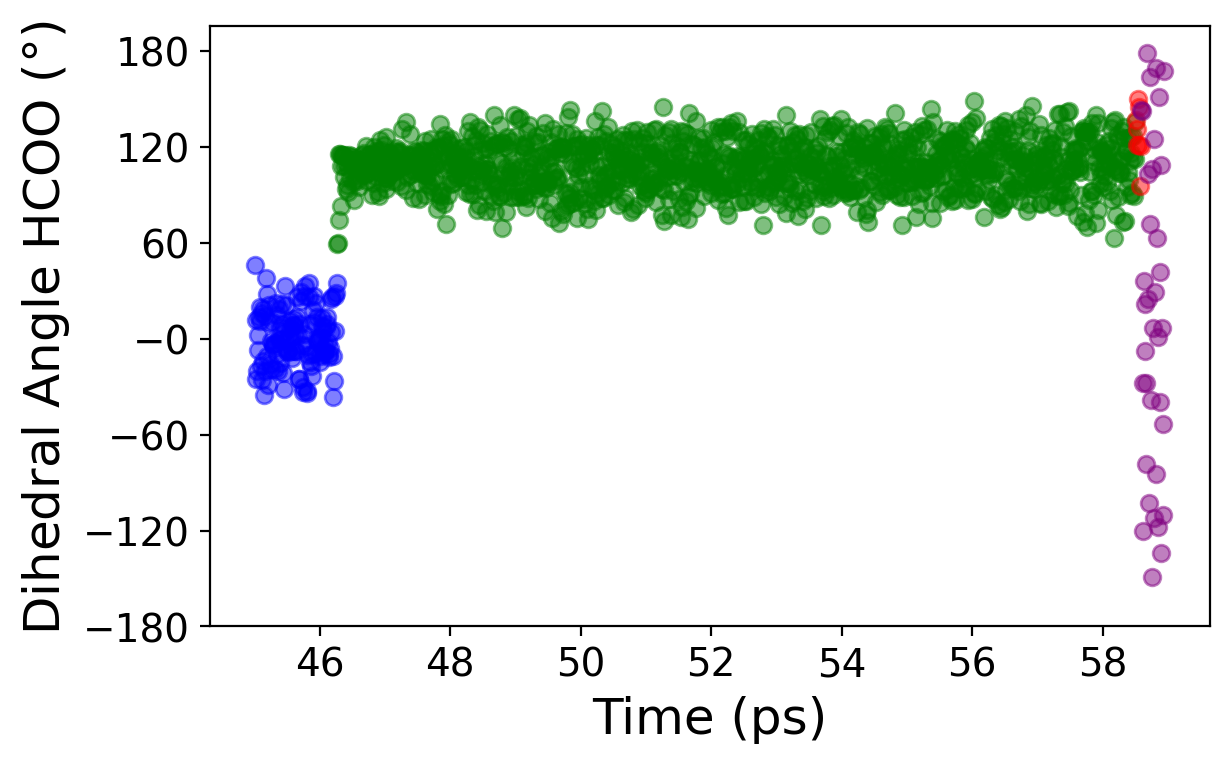}
    \caption{The change of dihedral angle HCOO along a typical
      reactive trajectory forming HCO+OH. Reactant H$_2$COO (blue)
      vibrates near the planar equilibrium structure ($0^{\circ}$),
      cyc-H$_2$CO$_2$ (green) stays near its equilibrium structure
      ($108^{\circ}$). For HCOOH (purple), two minima exist: {\it
        trans}-HCOOH (0$^{\circ}$) is 4 kcal/mol lower than {\it
        cis}-HCOOH ($\pm180^{\circ}$) and the TS between them is
      around 7 kcal/mol relative to {\it cis}-HCOOH.\cite{de2022fa}
      with over 100 kcal/mol energy, HCOOH can twist and deform
      freely, so the dihedral angle covers all the range.}
    \label{fig:dihe}
\end{figure}

\noindent
The HCO+OH channel is the one that was investigated in recent
photodissociation experiments.\cite{lester:2024} This product state
has been referred to as a ``minor channel''\cite{stanton:2015} whereas
reaction products CO$_2$+H$_2$ or H$_2$O+CO are expected to be
considerably more prevalent. Nevertheless, it is of interest to
analyze the dynamics that connects HCOOH and the HCO+OH
asymptote. Since HCOOH is highly energized ($\sim 140$ kcal/mol) large
amplitude motions within formic acid are excited. One of them is the
rotation of the two moieties, HCO and OH, relative to one
another. This is shown in Figure \ref{fig:dihe} which reports the HCOO
dihedral angle along a typical reactive trajectory. Given 25.5
kcal/mol initial energy, the reactant H$_2$COO (blue) vibrates near
its planar equilibrium structure at 0$^{\circ}$, while cyc-H$_2$CO$_2$
(green) remains close to its equilibrium dihedral angle $\sim
110^{\circ}$. For HCOOH (purple), there are two minima: the trans form
(0$^{\circ}$) is 4 kcal/mol more stable than the cis form
($\pm180^{\circ}$), with a TS between them about 7 kcal/mol above the
cis form.\cite{de2022fa} At energies exceeding 100 kcal/mol, HCOOH
undergoes extensive twisting and deformation, allowing its dihedral
angle to span the entire range.\\

\begin{figure} [H]
    \centering \includegraphics[width=0.8\linewidth]{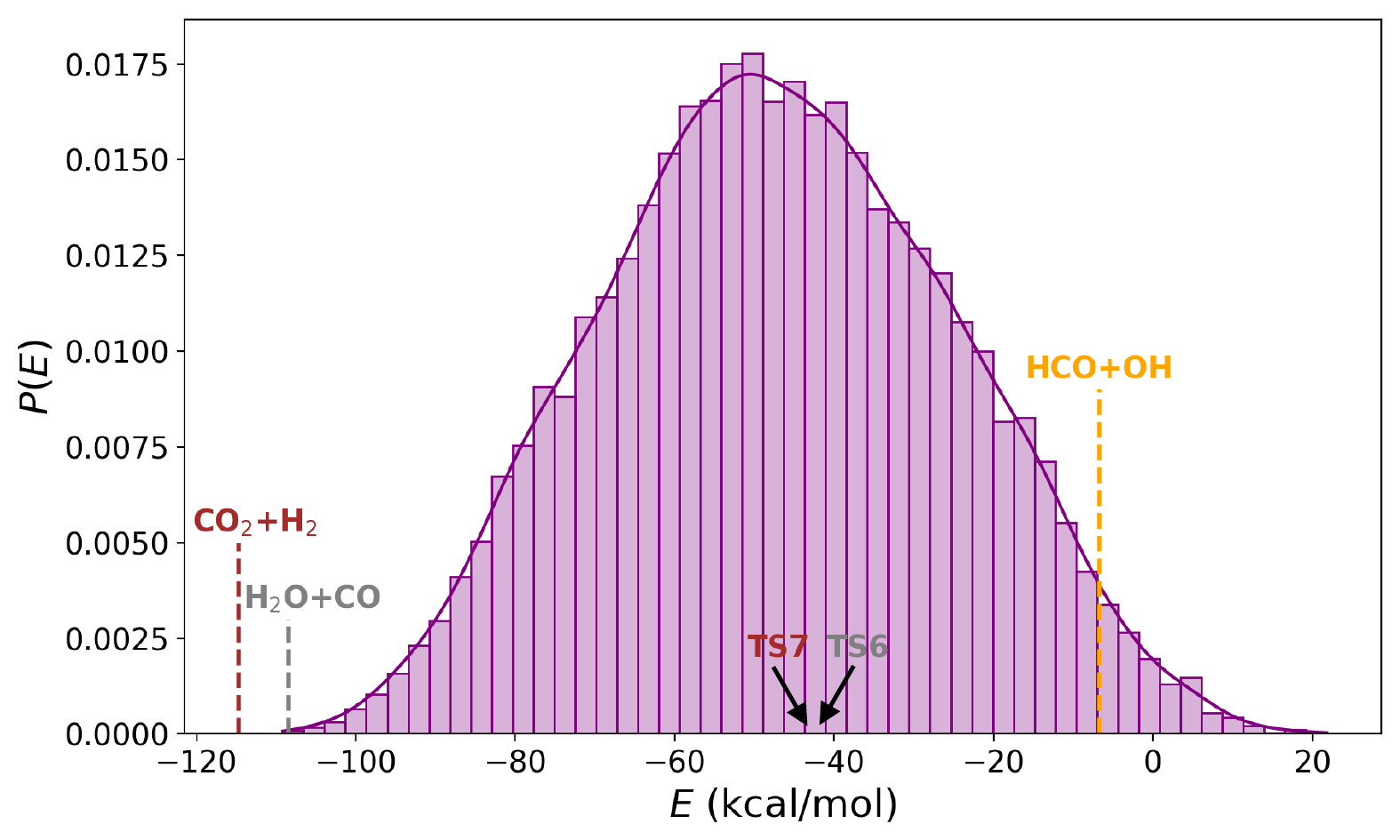}
    \caption{Energy distributions for samples of HCOOH from the
      simulation trajectories. The equilibrium energies of
      CO$_2$+H$_2$, H$_2$O+CO, and HCO+OH are denoted as black dashed
      lines, while TS6 and TS7 are indicated by arrows. Evidently,
      reaching energies to surmount TS6 and TS7 is facile, whereas the
      height of the HCO+OH asymptote is rarely sampled.}
    \label{fig:fa_en}
\end{figure}

\noindent
Finally, it is of interest to consider formation of the molecular
products in the context of the total energy stored within formic acid
after transiting TS4 (see Figure \ref{fig:diagram}). The internal
energy distribution is almost Gaussian and covers a range between
$\sim [-110,+20]$ kcal/mol, see Figure \ref{fig:fa_en}. The relative
energies of the diatomic products in Figure \ref{fig:fa_en} are shown
as dashed lines whereas transition states TS6 and TS7 are indicated by
arrows, see also Figure \ref{fig:diagram}. From this energy-centric
perspective it is predicted that only 3.1 \% of the trajectories
sampling the HCOOH-well are sufficiently energized to reach the HCO+OH
asymptote which is also consistent with a fraction of 1.6 \% from
analyzing the branching ratios. This confirms previous findings that
HCO+OH is a ``minority channel''.\cite{stanton:2015,lester:2024} The
CO distance distribution for these samples is shown in Figure
\ref{sifig:fa_co}.\\

\section{Discussion and Conclusion}
This work demonstrates that vibrational excitation of predominantly
the CH-stretch mode in H$_2$COO near the barrier TS1 to form dioxirane
launches non-equilibrium photodissociation dynamics to molecular
products including HCO+OH, CO$_2$+H$_2$, and H$_2$O+CO on the
nanosecond time scale. This is consistent with recent
simulations\cite{MM.h2coo:2024} and experiments.\cite{lester:2024} It
was also confirmed quantitatively, that the HCO+OH channel is a
minority channel (1.6 \%) as had been proposed some 10
years ago by computations\cite{stanton:2015} and 0.3 \% from a Master
equation calculation.\cite{stone:2018} The statistically significant
number of trajectories that was run and analyzed by using a
NN-based representation (here PhysNet) provides error bars on
fragmentation ratios and sufficiently sample conformational space to
characterize different possible reaction pathways. As an example, the
simulations found two competing routes between di-radical OCH$_2$O and
CO$_2$+H$_2$.\\

\noindent
To verify that excitation closer in energy to the transition state TS1
also leads to dioxirane, and subsequent reaction intermediates and
final products, additional simulations using different excitation
schemes and energies were carried out for a smaller number of
trajectories (100). First, the $2 \nu_{\mathrm{CH}} + 2
\nu_{\mathrm{COO}}$ combination mode was excited with 21 and 19
kcal/mol, respectively. Again, due to partial IVR, dwell times in the
reactant state were considerably longer: on a timescale of 1 and 5 ns,
diatomic product formation was observed in 17 and 77 out of 100
trajectories at 21 kcal/mol, and for 2 and 24 out of 100 trajectories
at 19 kcal/mol. Additional MD simulations with initial conditions
closest to the experiment were also performed. In this case only 2
quanta along the CH-stretch mode (18 kcal/mol, consistent with
experiments\cite{lester:2024}) without exciting the bending mode were
used for non-equilibrium excitation. Out of the 100 trajectories run
for 1 and 5 ns, 0 and 3 barrier crossings to dioxirane were found. In
other words, it was established that even for excitation close to the
TS1-barrier (20.55 kcal/mol for PES2025) progress towards molecular
products occurs.\\

\noindent
Experimentally, predominantly the $2 \nu_1$ mode in the region of the
CH-stretch overtone with an energy of 17 to 18 kcal/mol was
excited.\cite{lester:2024} The rotational temperature was reported to
be $T_{\rm rot} \sim 10$ K whereas for the vibrations, which are more
relevant and more difficult to cool, no specific information was
provided.\cite{lester:2024} The two methods for "in situ"
generation\cite{welz:2012,lester:2012} of H$_2$COO differ in
particular to which degree specifically high-frequency vibrations are
cooled.\cite{Hemberger:2022} Upon vibrational excitation any
population above the vibrational ground state can reach TS1 ``over the
barrier'' to form dioxirane whereas for the remainder of the
population reaching TS1 ($\approx 20$ kcal/mol above the reactant
state) is either energetically unfeasible or may require tunneling. A
comparatively slow reaction time and low OH-yield as found in the
present work (3 \% yield for excitation energies close to the TS1
barrier on the 5 ns time scale) is consistent with a measured OH-rise
rate of $\sim 10^6$ s$^{-1}$.\cite{lester:2024} On the other hand,
direct and converged ``brute force simulation'' of processes on the
$\mu$s time scale are still unfeasible even with a ML-PES. It should,
however, be stressed that ``exact'' reproduction of the experimental
conditions is neither possible nor necessary for drawing meaningful
conclusions from MD simulations but on considerably longer time
scales. Due to redistribution of the energy deposited along the
CH-stretch mode, in particular through coupling to the COO
angle,\cite{MM.h2coo:2024} and randomization on the $\sim 10$ ps time
scale (equivalent to hundreds of CO-vibrational periods) before
crossing TS1, the main findings of the present work are not expected
to be affected.\\

\noindent
Calculations following the small curvature semiclassical adiabatic
ground-state approach
(SCSAG)\cite{skodje1981general,skodje1982vibrationally} found that for
the first step along the low-energy pathway (1,3-ring closure) of
H$_2$COO the tunneling correction amounts to 1.8 and 2.5 at $T=300$ K
and 250 K, respectively.\cite{yin2017does} Such calculations were
based on the reaction path curvature from a B3LYP/6-311+G(2d,2p) IRC
calculations with energies recalculated at the QCISD(T) complete basis
set limit.\cite{peterson1994benchmark} Corrections of this magnitude
indicate that tunneling may contribute to the rate but that this is
not the prime factor. Unfortunately, the tunneling corrections from a
combined equilibrium RRKM treatment with three different types of
tunneling calculations (1D-asymmetric Eckart barrier,
1d-Wentzel-Kramers-Brillouin calculation, and
multidimensional-semiclassical transition state theory (SCTST)) were
not explicitly reported\cite{lester:2024} which prevents direct
comparison with the tunneling corrections found from SCSAG. Rather, it
was concluded that the reaction must occur in a classically forbidden
regime because the initial excitation energy lies below the computed
barrier which, however, is rather indirect and needs to assume that
the computed barrier is sufficiently accurate. On the other hand,
previous experiments reported a barrier height of $19.55 \pm 1.48$
kcal/mol for TS1 which covers a range [18.1,21.0] kcal/mol by matching
experimental and computed rates from a Master equation
analysis.\cite{stone:2018} It is noted that the lower end of the
interval (18.1 kcal/mol) is close to the estimated excitation energy
used in recent experiments.\cite{lester:2024}\\

\noindent
As a comparison between the two different types of tunneling
calculations the next-larger CI, CH$_3$CHOO, is considered. Tunneling
corrections for the HT reaction pathway determined from SCSAG
calculations\cite{yin2017does} at $T = 300$ and 250 K were 70 and
2000, respectively. These values compare with corrections of $\sim
500$ ($T = 300$ K) and 4000 ($T = 250$ K) from SCTST tunneling
calculations\cite{fang:2016deep} which establishes the validity of the
SCSAG approach. In other words, a factor of 1.8 to 2.5 for the
tunneling correction for H$_2$COO as found from SCSAG calculations
indicate that tunneling may contribute but is not the only factor. On
the other hand, because the classical MD simulations neglect the
effect of tunneling on the reaction rates, the kinetic traces for
product yield should be considered as lower limits to the true rates
for molecular product formation expected from future experiments.\\

\noindent
An interesting and potentially relevant result for atmospheric
chemistry is the finding that none of the molecular intermediates -
dioxirane, di-radical OCH$_2$O, and HCOOH (formic acid) -
accumulate. Once the reaction is launched past TS1 towards dioxirane,
progress to molecular products (HCO+OH, CO$_2$+H$_2$, and H$_2$O+CO)
occurs on the picosecond time scale. CIs are expected to be found in
the troposphere and lower stratosphere\cite{johnson:2008} where
collision times with other molecular species occur on the $\mu$s to ns
(troposphere) and on the ms to $\mu$s-time scale (lower
stratosphere). Under such conditions it is conceivable that
collisional stabilization for reaction intermediates occur. Such
processes are clearly not included in the present work and require
dedicated work based on the present computational models. The
quenching of vibrational excitations due to IVR, which occurs
primarily through coupling to other internal degrees of freedom (in
particular the COO-bend, found in the present case has already been
found for other compounds relevant to atmospheric chemistry, such as
acetaldehyde.\cite{MM.atmos:2020}\\

\noindent
In conclusion, the present work provides the first end-to-end
characterization of the vibrationally induced dissociation dynamics of
H$_2$COO to molecular products. It is demonstrated that NN-based
reactive PESs are instrumental to such studies and provide the
necessary computational framework for quantitative studies. This opens
the door for wider exploration of reaction types and mechanisms from
computer-based models to provide molecular-level insight and
understanding of rather complex processes relevant to the
atmosphere.\\

\section*{Supporting Information} 
The supporting information reports the CASPT2 energies for the HCO+OH
asymptote, the energy profile for the IRC from H$_2$COO to
cyc-H$_2$CO$_2$, the correlation plot of the 5162 original samplings
between PES2025 and PES2024, the CH bond changes over time, the
time-dependent population changes of each species in a 5 ns
simulation, the reaction coordinate as a function of time for the
three product channels, the CO distance distributions for samples of
HCOOH, and the geometrical criteria used to assign structures along
the trajectories.\\

\section*{Data Availability} 
The reference data that allow to reproduce the findings of this study
are openly available at \url{https://github.com/MMunibas/h2coo.full}.\\

\section*{Acknowledgment}
Financial support from the Swiss National Science Foundation through
grants $200020\_219779$ (MM), $200021\_215088$ (MM), the University of
Basel (MM) is gratefully acknowledged. This article is based upon work
within COST Action COSY CA21101, supported by COST (European
Cooperation in Science and Technology) (to MM). The authors thank
D. L. Osborn, M. I. Lester and M. Suhm for correspondence.

\bibliography{refs.clean}

\clearpage

\renewcommand{\thetable}{S\arabic{table}}
\renewcommand{\thefigure}{S\arabic{figure}}
\renewcommand{\thesection}{S\arabic{section}}
\renewcommand{\d}{\text{d}}
\setcounter{figure}{0}  
\setcounter{section}{0}  
\setcounter{table}{0}

\newpage

\noindent
{\bf SUPPORTING INFORMATION: End-to-End Photodissociation Dynamics of
  Energized H$_2$COO}

\begin{figure} [H]
    \centering
    \includegraphics[width=0.8\linewidth]{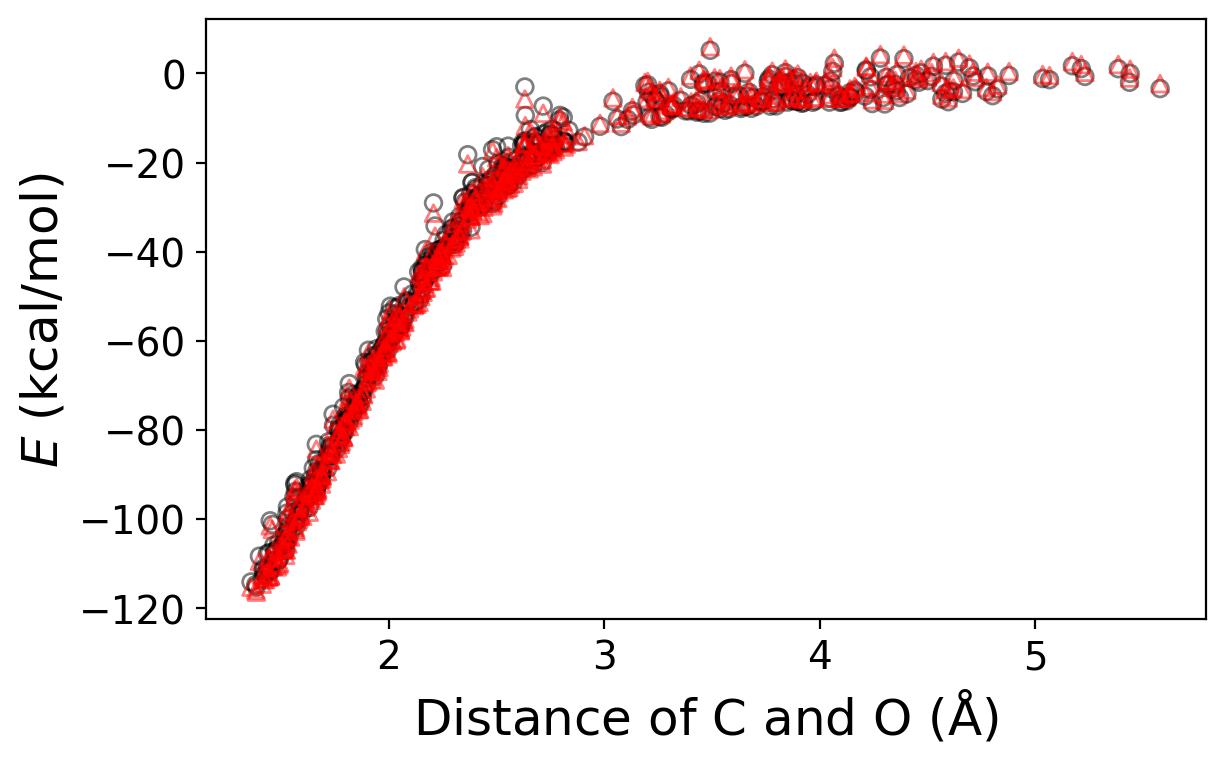}
    \caption{The sampled CASPT2(black circle)/PES2025(red triangle)
      energies for the HCO$\cdot$+$\cdot$OH asymptote were obtained
      through a relaxation scan of the distance between HCO and OH,
      followed by random displacements of up to 0.3 \AA\/ for each
      optimized geometry. Starting from $\sim 1800$ geometries, only
      for about one third ($\sim 600$) the CASPT2 calculations
      converged. Despite this, these samples provide sufficient
      information to train meaningful PhsNet models towards HCO+OH
      dissociation. The MAE/RMSE between the reference and PES2025
      energies are 0.99/1.13 kcal/mol and $R^2 = 0.9995$.}
    \label{sifig:hco_oh_cas}
\end{figure}

\begin{figure} [H]
    \centering
    \includegraphics[width=0.8\linewidth]{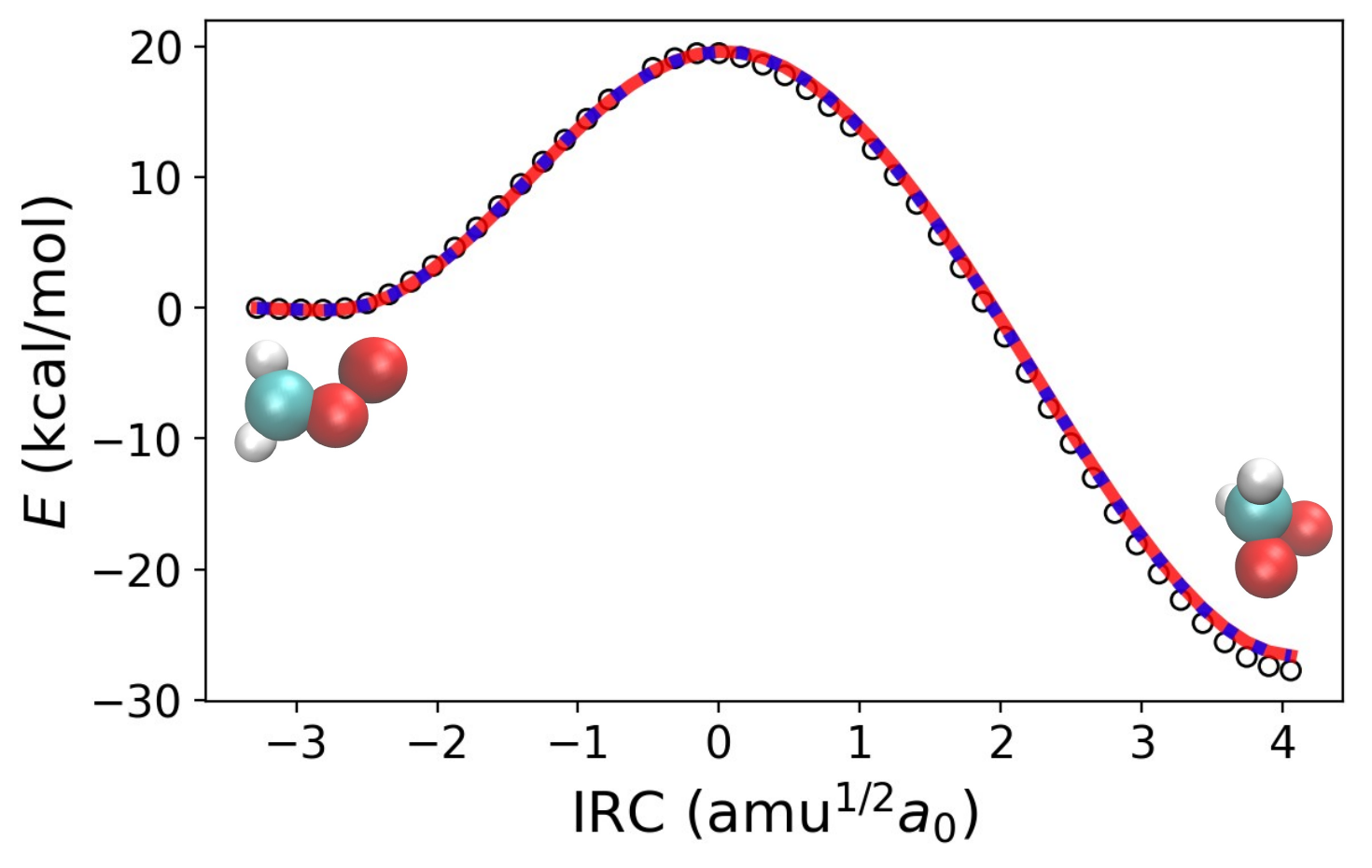}
    \caption{The energy profile for the IRC between H$_2$COO and
      cyc-H$_2$CO$_2$ across TS1. The geometries were obtained from an
      IRC calculation at the MP2/aVTZ level. The black circles
      represent the CASPT2/aVTZ data, the red line corresponds to the
      PES2025 surface, and the blue dotted line indicates the PES2024
      surface.}
     \label{sifig:irc1}
\end{figure}

\begin{figure} [H]
    \centering \includegraphics[width=0.8\linewidth]{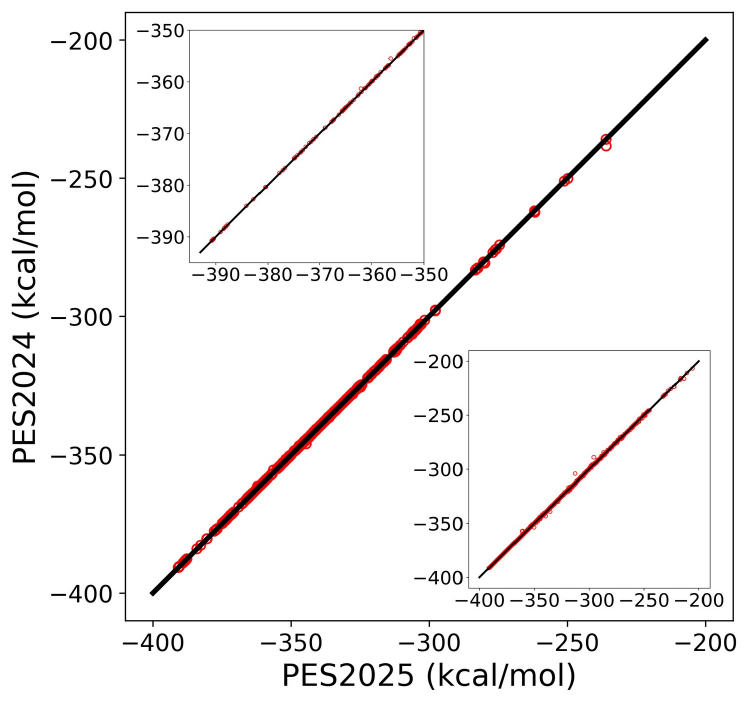}
    \caption{The
      correlation plot of the 5162 original samplings between PES2025
      and PES2024 is shown. The main panel shows the test dataset, with a magnified view in the upper-left inset, while the bottom-left inset displays the training dataset. The MAE, RMSE, and $R^2$ values are 0.06/0.07 kcal/mol,
      0.17/0.23 kcal/mol, and 0.999949/0.999907 for the test/train
      datasets, respectively.}
    \label{sifig:2pes}
\end{figure}

\begin{figure} [H]
    \centering
    \includegraphics[width=0.8\linewidth]{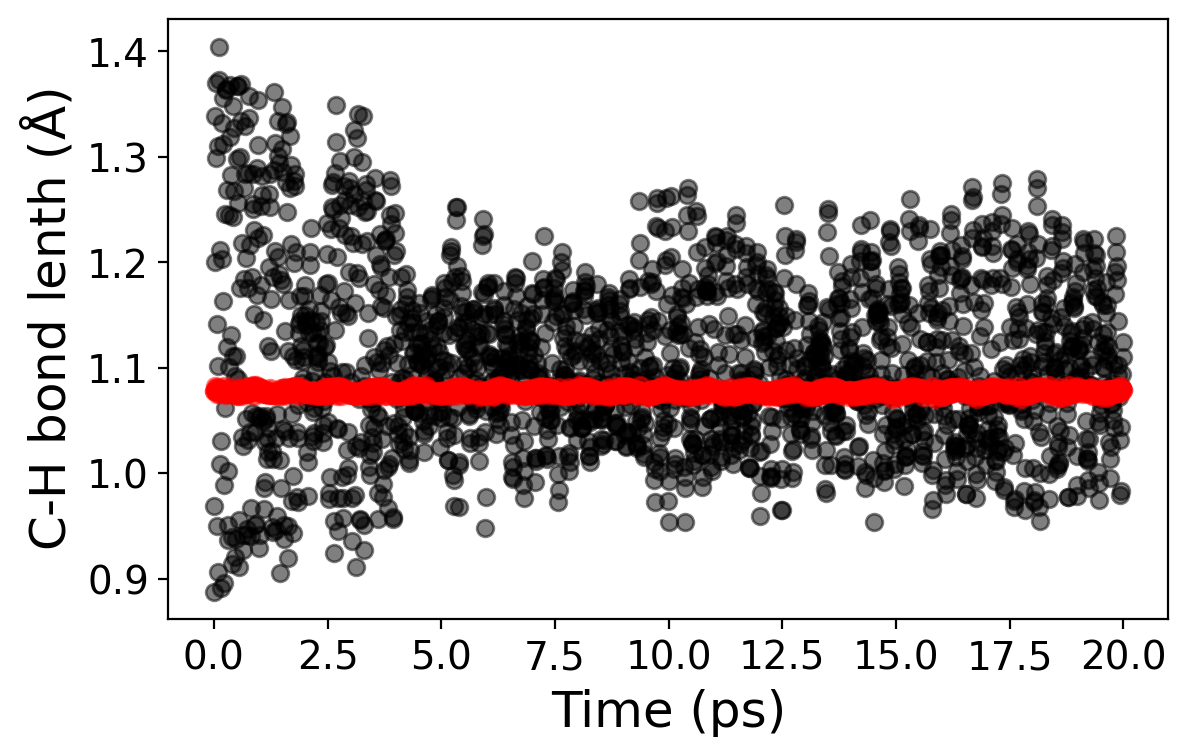}
    \caption{The excitation energy was added to the CH local mode with
      3 quanta by scaling the velocities of the carbon atom and the
      hydrogen atom closer to the terminal oxygen, in order to match
      24 kcal/mol of energy. Additionally, the COO bending mode was
      excited with 1 quantum (1.5 kcal/mol) of energy. The change in
      the CH bond length can serve as an important indicator of
      Intramolecular Vibrational Energy Redistribution (IVR) since
      most of the excitation energy was added to the CH mode. Red
      points are from equilibrium sampling at 300K. Partial IVR occurs
      within the first $\sim 4$ ps, during which the energy initially
      in the CH-stretching mode is redistributed into other
      vibrational modes. The reaction takes place after the IVR
      process. Among the $\sim 2800$ reactive trajectories, the first
      occurrence of
      cyc-H$_2$CO$_2$/CO$_2$+H$_2$/HCOOH/H$_2$O+CO/HCO$\cdot$+$\cdot$OH
      was observed at 9.8/11.8/16.1/21.0/26.6 ps, respectively.}
    \label{sifig:ch}
\end{figure}

\begin{figure} [H]
    \centering
    \includegraphics[width=0.8\linewidth]{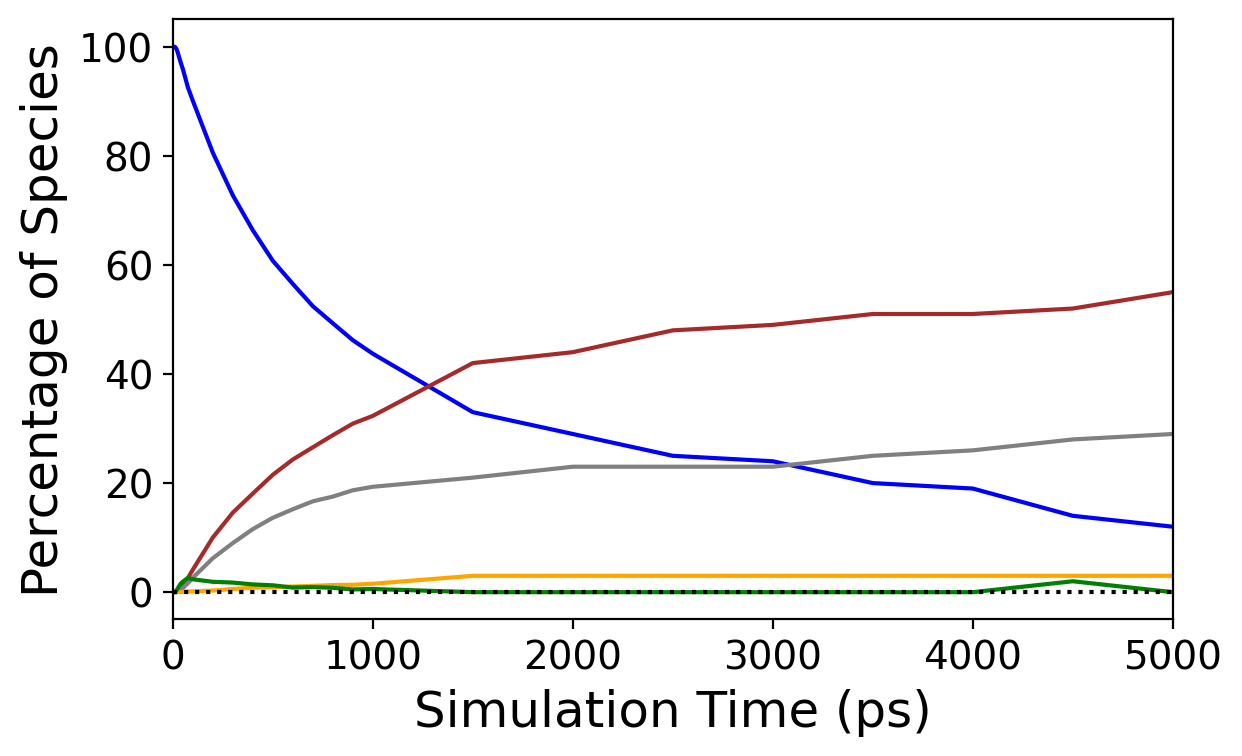}
    \caption{Changes in time-dependent populations for H$_2$COO
      (blue), cyc-H$_2$CO$_2$ (green), HCO$\cdot$+$\cdot$OH (Orange)
      CO$_2$+H$_2$ (brown), and H$_2$O+CO (gray) during a 5 ns
      simulation. Note that after 1 ns, only 100 trajectories were
      performed.}
    \label{sifig:popu_5000}
\end{figure}

\begin{figure} [H]
    \centering \includegraphics[width=0.8\linewidth]{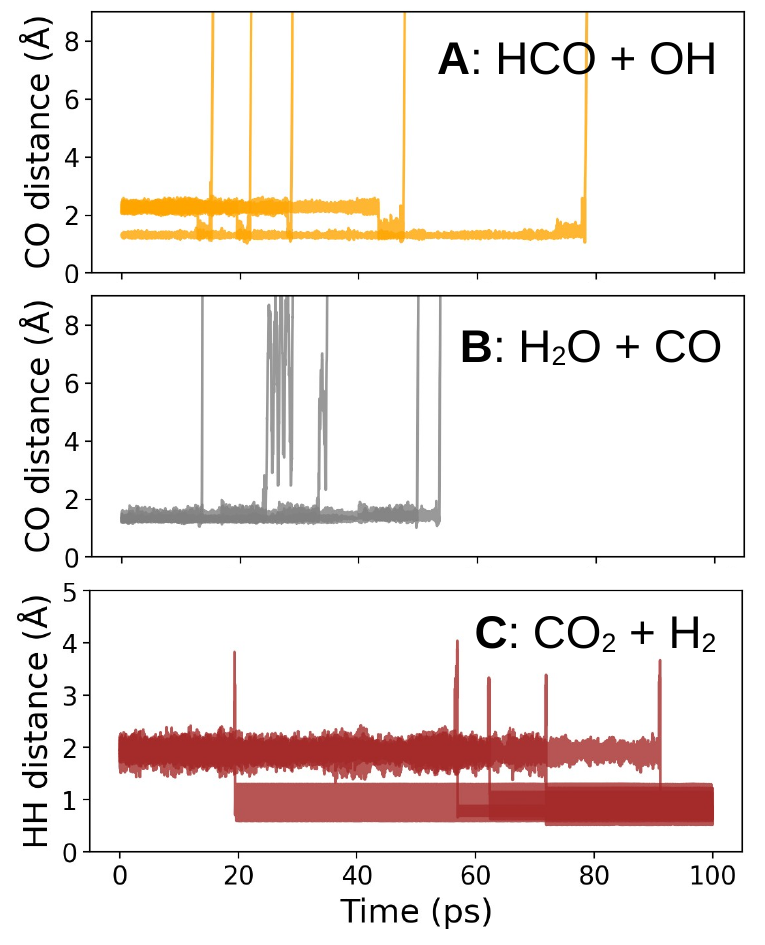}
    \caption{Reaction coordinate as a function of time for the three
      product channels. As soon as sufficient energy is available
      along the reaction coordinate, dissociation can occur. Reaction
      coordinates are the C$_{\rm HCO}$--O$_{\rm OH}$, O$_{\rm
        H_2O}$--C$_{\rm CO}$, and H--H separations.}
    \label{sifig:react-coor}
\end{figure}

\begin{figure} [H]
    \centering \includegraphics[width=0.8\linewidth]{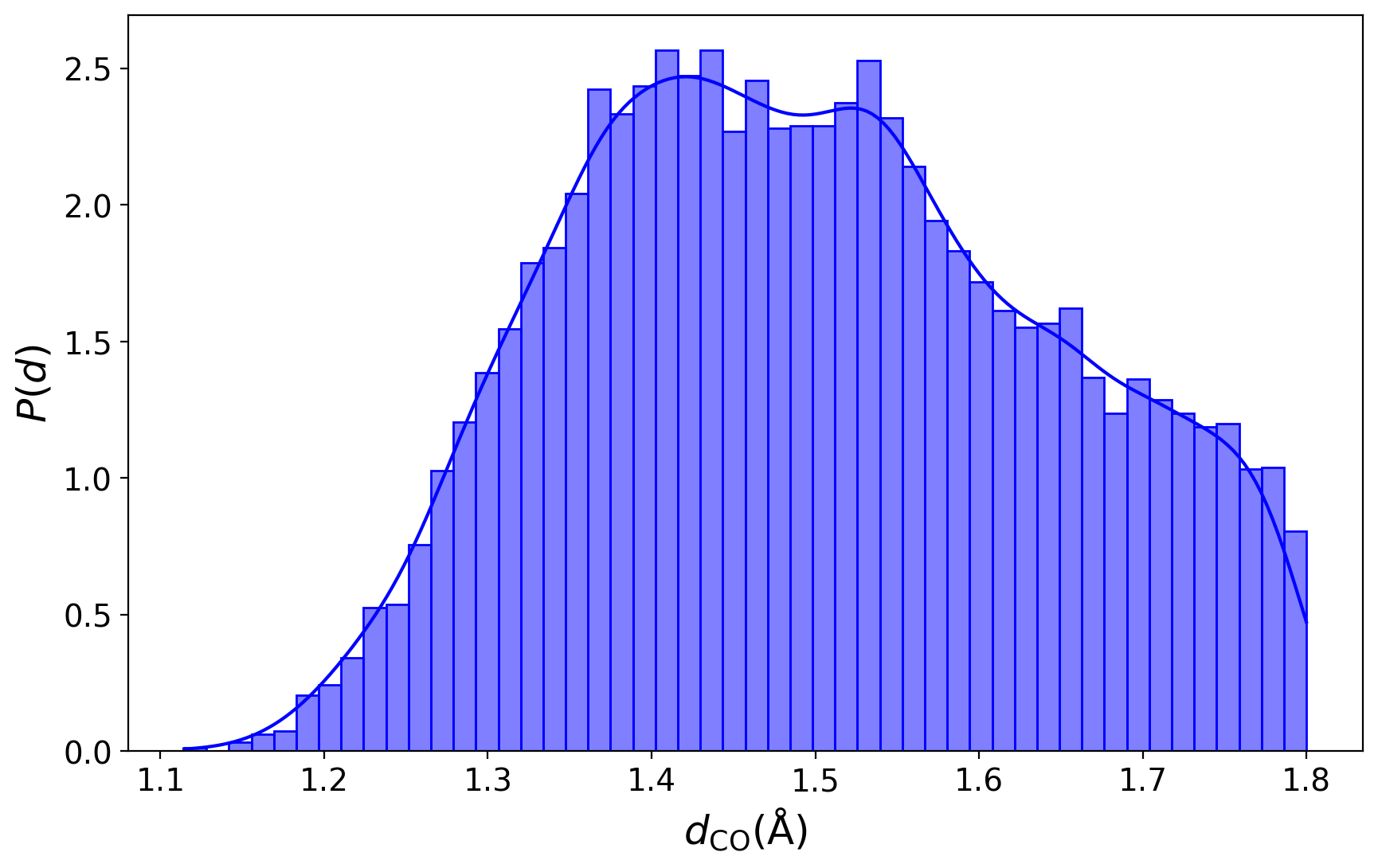}
    \caption{CO distance distributions for samples of HCOOH from the
      simulation trajectories are presented. The cutoff at 1.8 \AA\ is
      based on the criteria outlined in Table \ref{sitab:criteria}, to
      distinguish HCOOH from HCO+OH.}
    \label{sifig:fa_co}
\end{figure}

\begin{table} [H]
    \centering
    \begin{tabular}{c|c|c|c|c}
    \hline
    Species & H$_2$COO & cyc-H$_2$CO$_2$ & OCH$_2$O & HCOOH \\
    \hline
    distance CH & $<$ 1.8 & $<$ 1.5 & $<$ 1.5 & $<$ 1.5, $>$ 1.5 \\
    \hline
    distance CO & $<$ 1.8, $>$ 1.7 & $<$ 2.0 & $<$ 2.0 & $<$ 1.6, $<$ 1.8 \\
    \hline
    distance OO & $<$ 2.3 & n.u. & n.u. & $>$ 2.0 \\
    \hline
    angle COO & $>$ 90, $<$ 90 & $<$ 90 & $<$ 90 & $<$ 90 \\
    \hline
    angle OCO & $<$ 90 & $<$ 95 & $\geq$ 95 & $>$ 90 \\
    \hline
    \end{tabular}
    \caption{Geometrical criteria were used to assign structures along
      the trajectories. The units of distance and angle are \AA/ and
      $^{\circ}$, respectively. For the two CH and CO bond lengths and
      the two possible COO angles, the criteria were applied twice
      when necessary.``n.u.'' stands for ``not used''.}
    \label{sitab:criteria}
\end{table}

\end{document}